\documentclass[10pt,twocolumn,letterpaper]{article}
\usepackage{iccv}

\usepackage{times}
\usepackage{graphicx}
\usepackage{amsmath}
\usepackage{amssymb}
\usepackage{subcaption}
\usepackage{enumitem}
\usepackage{appendix}
\setlist{leftmargin=*}

\newcommand{\bfx}{{\bf x}}

\newcommand{\bfw}{{\bf w}}

\usepackage[pagebackref=true,breaklinks=true,letterpaper=true,colorlinks,bookmarks=false]{hyperref}

\iccvfinalcopy 

\def\iccvPaperID{104} 

\ificcvfinal\fi
\begin{document}

\title{Programmable Spectrometry --- Per-pixel Classification of Materials using Learned Spectral Filters}

\author{Vishwanath Saragadam, and Aswin C.~ Sankaranarayanan\\
Department of ECE, Carnegie Mellon University, USA\\
{\tt\small vishwanathsrv@cmu.edu}
}

\maketitle

\begin{abstract}
Many materials have distinct spectral profiles.
This facilitates estimation of the material composition of a scene at each pixel by first acquiring its hyperspectral image, and subsequently filtering it using a bank of spectral profiles.
This process is  inherently wasteful since only a set of linear projections of the acquired measurements contribute to the classification task.
We propose a novel programmable camera that is capable of producing images of a scene with an arbitrary spectral filter.
We use this camera to optically implement the spectral filtering of the scene's hyperspectral image with the bank of spectral profiles needed to perform per-pixel material classification.
%
This provides gains both in terms of acquisition speed --- since only the relevant measurements are acquired --- and in signal-to-noise ratio --- since we invariably avoid narrowband filters that are light inefficient.
Given training data, we use a range of classical and modern techniques including SVMs and neural networks to identify the bank of spectral profiles that facilitate material classification.
We verify the method in simulations on standard datasets as well as real data using a lab prototype of the camera.
%
%
%
%
%
\end{abstract}

\section{Introduction}\label{section:intro}
Material composition of a scene can often be identified by analyzing variations of light intensity as a function of spectrum or wavelengths.
%
Since materials tend to have unique spectral profiles, spectrum-based material classification has found widespread use in numerous scientific disciplines including molecular identification using Raman spectroscopy \cite{colthup2012introduction}, tagging of key cellular components in fluorescence microscopy \cite{lichtman2005fluorescence}, land coverage and weather monitoring \cite{cloutis1996review,harsanyi1994hyperspectral}, and even the study of chemical composition of stars and astronomical objects using line spectroscopy.
It would not be a stretch to suggest that spectroscopy or its imaging variant, hyperspectral imaging (HSI), is an important scientific tool for material identification.
\begin{figure}[!tt]
	\centering
	\includegraphics[width=0.95\columnwidth]{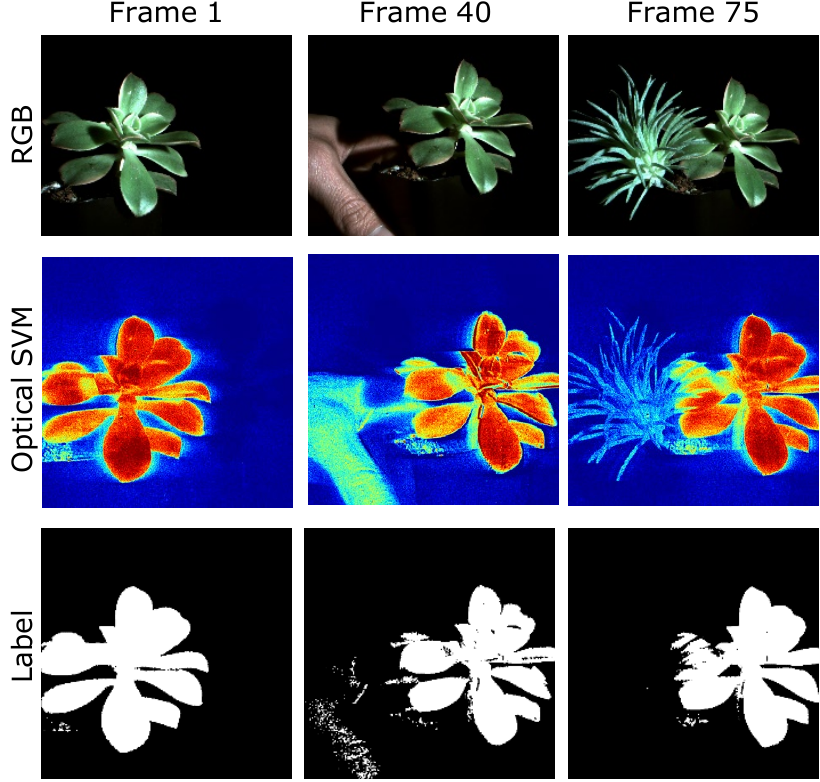}
	\caption{\small\textbf{Video-rate material classifier.} We propose an optical setup that is capable of classifying material on a per-pixel basis. This is achieved by building a programmable spectral filter that can image at high spatial resolution. The images here show a video sequence of a identifying real plants from plastic plants. We captured the data at $4$fps, and performed only a per-pixel thresholding to get the video result.}
	\label{fig:teaser}
\end{figure}

While hyperspectral imaging has also found application in computer vision tasks \cite{kim20123d,pan2003face,tarabalka2010segmentation}, its widespread adoption  has been hindered due to inherent challenges in acquisition them.
Capturing a HSI requires sampling of a very high dimensional signal; for example, mega-pixel images at hundreds of spectral bands, a process that is daunting to do at video rate.
This problem is further aggravated by the fact that hyperspectral measurements have to combat low signal to noise ratios, as a fixed amount of light is divided in to several spectral bands --- leading to long exposure times that can even span  several minutes per HSI.
%

This paper proposes a novel approach for enabling spectrometry-based  per-pixel material classification by overcoming the limitations posed by HSI acquisition.
To understand our proposed approach, we first need to delve deeper into the process of classification itself.
Classification  techniques involve comparing the spectral profile at each pixel with known or \emph{learned} spectra by taking a linear projection.
Intuitively, given $K$ material classes, we would compute $\mathcal{O}(K)$ such linear projections. 
For example, a support vector machine (SVM) classifies by finding distance of features from the separating hyperplane; in the context of spectral classification, this translates to spectrally filtering the scene with the hyperplane coefficients.
Hence, spectral classification can be made practical if we can capture the linear projections directly without having to acquire the complete HSI.
Such an operation translates to optically filtering the scene's HSI using known spectral filters, which can be achieved if the camera's spectral response can be arbitrarily programmed.

To enable per-pixel material classification, we propose a new imaging architecture with a programmable spectral response that can changed on-the-fly at video rate.
Given a training dataset of spectral profiles, we use off-the-shelf classification techniques like SVMs and deep neural networks to identify linear projections that facilitate material classification.
For a novel scene, the camera captures multiple images, each with a different spectral response; the captured measurements are used  with the classifier to perform per-pixel material classification.
%
%
%
%
%

The proposed pipeline has numerous benefits.
\textit{Optical computing} of the linear projections allows us to circumvent the measurement of sampling the full HSI.
This has the dual benefit of reducing the acquisition time (from minutes to hundreds of milliseconds) as well as increasing light efficiency of each captured image since the linear projections often correspond to broadband spectral profiles.
For binary classification problem, our lab prototype provides a classification result every second frame thereby providing material labels at 4 frames per second.
We also show results on multi-class labeling problems using a classifier that can differentiate between five distinct material types.
%
%
%

\section{Prior Work}\label{section:prior}

We discuss prior work in the areas of material classification using HSIs as well as optical computing and design of programmable spectral filters.

\paragraph{Hyperspectral classification.}
Consider the HSI of a scene, $H(x, y, \lambda)$, where each pixel $(x, y)$ is assumed to belong to one of  $K$ material classes. 
Specifically, the spectra at each pixel can be written as,
\begin{align}
	H(x, y, \lambda) = \alpha (x, y) S_{L(x, y)} (\lambda),
	\label{eq:model}
\end{align}
where $L(x, y)$ is label of the material contributing to spectrum at $(x, y)$, and $\alpha(x, y)$ is scaling parameter.
Note that the model above assumes all spatial pixels are pure, i.e., every pixel gets contribution from only one material.
%
We use this model for the sake of exposition and later discuss about how to relax it later to handle mixed pixel.

The goal of classification is to estimate the label at each pixel, $L(x, y)$, which forms a label map.
There are broadly two approaches to spectral classification --- generative and discriminative.
Generative techniques rely on decomposing the HSI as a linear combination of basic materials that are called end-members \cite{dobigeon2014nonlinear}.
Specifically, the HSI of the scene is decomposed as,
\begin{align}
	H(x, y, \lambda) &= \sum_{k=1}^{K} s_k(\lambda ) a_k(x, y),
\end{align}
where $s_k(\lambda)$ is the spectra of $k^\text{th}$ material, and $a(x, y)$ is the relative contribution of material $k$ at $(x, y)$.
The abundances at each pixel along with the end-member spectra provide a feature vector that can be used to spatially cluster the materials and subsequently identify them.

Discriminative techniques rely on directly learning discerning features from the HSI without the intermittent stage of low-dimensional decomposition.
Here, we identify  a set of spectral filters, $\left\{ (d_k(\lambda), \beta_k) \right\}_{k=1}^{M}$ that generate per-pixel feature vector via spectral-domain filtering:
\begin{align}
	F_k(x, y) = \int_{\lambda} H(x, y, \lambda) d_k (\lambda) d\lambda + \beta_k.
	\label{eq:discriminating}
\end{align}
Hence, each image $F_k(x, y)$ is a spectrally-filtered version of the HSI with an added offset. 
%
In case of SVMs, the learned spectral filters form separating hyperplanes; this has been a \textit{de facto} way of HSI classification \cite{fauvel2007spectral,melgani2004classification}.
More sophisticated learning techniques based on neural networks use spectral features  \cite{hu2015deep} or spatio-spectral features \cite{sharma2016hyperspectral,liu2017semi,hamida20183,lee2016contextual,chen2016deep,li2017spectral,luo2018hsi,he2017multi} for classification.
%
%
%
Invariably, the number of spectral features used, i.e, the dimensionality of the projection,  tends to be smaller than the number of spectral channels in the HSI.
%
%
Hence, we seek to measure the features directly, by   computing (\ref{eq:discriminating}) optically.
As is to expected, such a paradigm of \emph{optical classification} requires the design of cameras that can be programmed with arbitrary spectral filters.

\paragraph{Optical computing.}
Instead of relying on both spatial and spectral information, we consider a simpler approach which relies only on the spectral profiles for classification.
Such a strategy is less accurate than spatial and spectral versions \cite{sharma2016hyperspectral,liu2017semi,hamida20183,lee2016contextual,chen2016deep,li2017spectral,luo2018hsi,he2017multi}, but significantly reduces the complexity of the imaging system.
%
%
%
This approach is similar, in spirit, to using BRDFs  to perform per-pixel classification by varying the incident illumination \cite{liu2014discriminative,gross2002fisher}, or using first layer of a neural network to capture light fields \cite{chen2016asp}.
%
%
Such a setup offers two-fold advantage:
\begin{enumerate}
	\item \emph{Fewer measurements.} Since the number of material classes is far fewer than number of spectral bands, we need to measure far fewer measurements. For example, we show in our experiments that $3 - 5$ images suffice for a 5-class classification task.
	\item \emph{Increased SNR.} The discriminating filters tend to be spectrally broadband, and hence each image is measured at higher light levels than any individual narrow spectral band. Hence, the images can be captured at higher SNR or at faster acquisition rates.
\end{enumerate}
Optical computing has found use in various computer vision tasks such as capturing light transport matrices \cite{o2010optical}, low-rank approximation of hyperspectral images \cite{saragadam2018krism}, and spectral classification using programmable light sources \cite{goel2015hypercam,park2007multispectral}.
We adopt the paradigm of optical computing to make discriminative filter measurements by building a camera whose spectral response can be arbitrarily programmed.

\paragraph{Dynamic spectral filters.}
Spectral filtering can be achieved by modified the response of the camera; a canonical and static example being the Bayer pattern or more interestingly, the case of fluorescence filters in microscopy.
It is however more useful to have a camera whose response can be altered arbitrarily in a fast manner.
Numerous techniques to achieve spectral filtering have been proposed in the past.
Agile spectral imager \cite{mohan2008agile} rely on the coding the so-called ``rainbow plane" to achieve arbitrary spectral filtering. 
%
%
This was further developed by \cite{love2014full} where they placed a digital micromirror device (DMD) on the rainbow plane to achieve dynamic spectral filtering.

However, such architectures come with a debilitating problem --- usage of simple pupil codes such as open aperture or a slit directly tradeoff spatial resolution for spectral resolution.
This was first identified in \cite{saragadam2018krism} in the context of hyperspectral imaging. 
They showed that a slit, a common choice for spectrometry, leads to large spatial blur. 
Similarly an open aperture, a common choice for high-resolution imaging, leads to large spectral blur.
Hence, such apertures are  not capable of spectral classification with high accuracy.

We instead rely on the optical setup in \cite{saragadam2018krism} to overcome the spatial-spectral tradeoff.
The key idea is to use a coded aperture that introduces an invertible blur in both spatial and spectral domains.
An important difference is that the setup in \cite{saragadam2018krism} is designed for HSI image acquisition; this paper adapts the underlying ideas for performing material classification in the scene.

\section{Programmable Spectral Filter}
\label{section:coded}

\begin{figure}[!tt]
	\centering
	\begin{subfigure}[t]{\columnwidth}
		\centering
		\includegraphics[width=\columnwidth]{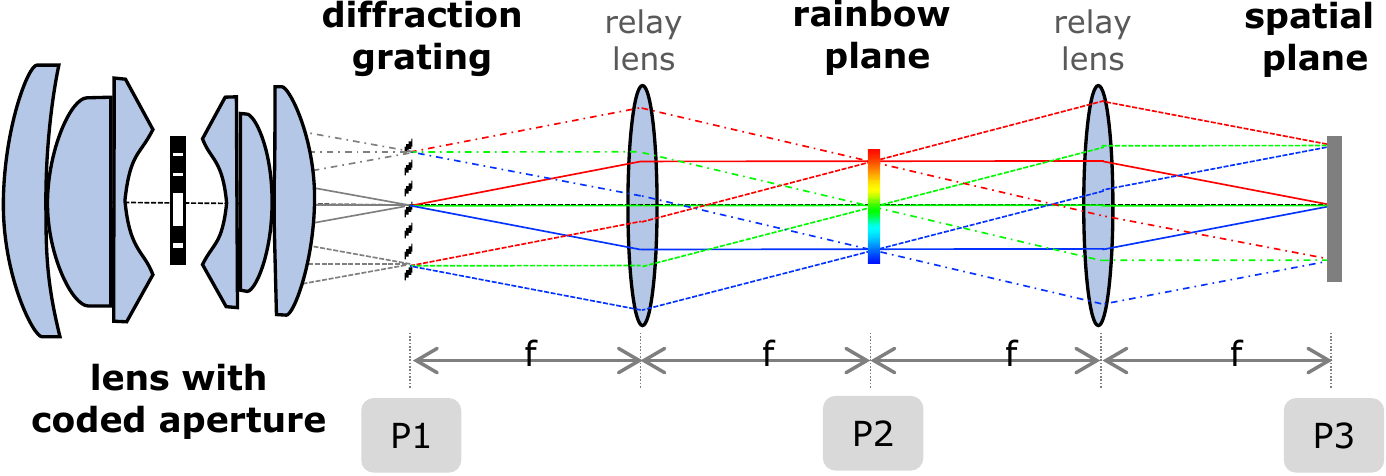}
		\caption{Setup schematic}
	\end{subfigure}
	\\
	\begin{subfigure}[t]{\columnwidth}
		\centering
		\includegraphics[width=\columnwidth]{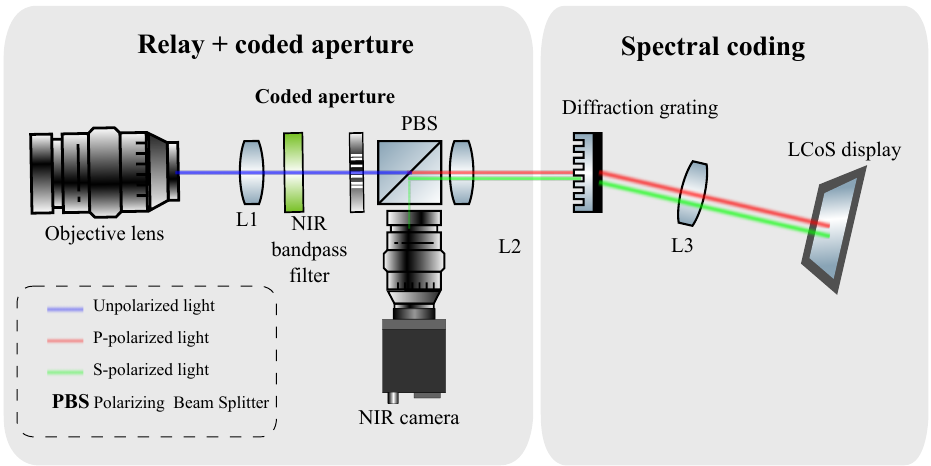}
		\caption{Practical realization}
	\end{subfigure}
	\caption{\small\textbf{Schematic for programmable spectral filter.} The optical architecture in (a) consists of a lens assembly with coded aperture which introduces spatial and spectral blurs. By placing an SLM in P2, the HSI of the scene can be spectrally filtered and sensed by a camera sensor on P3. (b) shows a compact realization of the optical setup.}
	\label{fig:4f}
\end{figure}

\begin{figure}[!tt]
	\centering
	\includegraphics[width=\columnwidth]{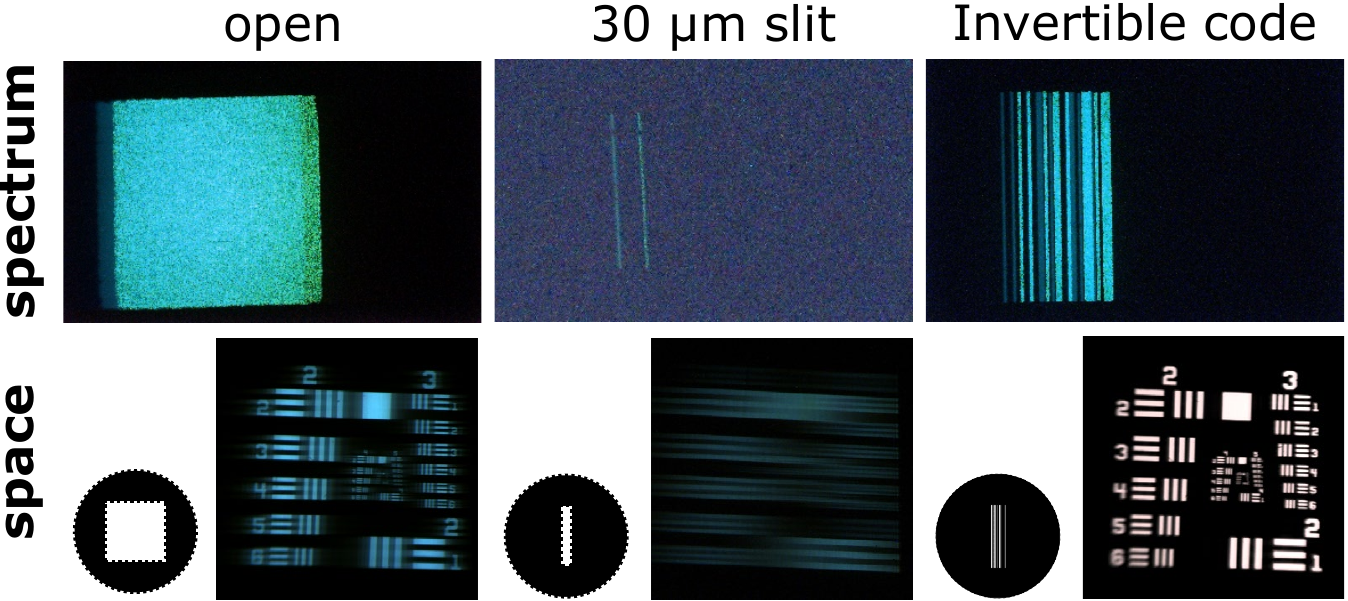}
	\caption{\small\textbf{Spatio-spectral resolution tradeoff.} A slit is capable of high spectral resolution whereas an open aperture is capable of high spatial resolution but both are inappropriate for high spatio-spectral HSI imaging. In contrast, a coded aperture introduces an invertible spatial and spectral blurs which can then be deconvolved. Figure reproduced with permission from \cite{saragadam2018krism}.}
	\label{fig:space_spectrum}
\end{figure}

Our optical setup is a modification of the optical setup proposed in \cite{saragadam2018krism}.
We briefly explain the relevant parts of the optical setup here. The interested reader is referred to \cite{saragadam2018krism} as well as appendix for a detailed derivation.
\paragraph{4f system for spectral programming.} We borrow the optical schematic for spectral programming from \cite{saragadam2018krism}, shown in Fig.\ \ref{fig:4f}(a).
%
%
Given the HSI, $H(x, y, \lambda)$, that is focused on the grating at P1, we seek to derive the intensity on planes P2 and P3.
%
%
The intensity on rainbow plane P2,
\begin{align}
	I_4(x, y) &= a^2(-x, -y) \ast \left(S\left(\frac{x}{f\nu_0}\right)\widetilde{c}\left(\frac{x}{f\nu_0}\right)\right),
	\label{eq:rainbow}
\end{align}
where $S(\lambda) = \int_{(x, y)}H(x, y, \lambda)$ is spectrum of the scene, $\widetilde{c}(\lambda)$ is response of the optical system, and $\nu_0$ is the density of groves in $mm^{-1}$.
The intensity on image plane P3,
\begin{align}
	I_5(x, y) &= \int_{\lambda} \left(H(x, y, \lambda ) \ast \left| \frac{1}{\lambda^2 f^2} A\left(-\frac{x}{\lambda f}, -\frac{y}{\lambda f}\right)\right|^2 \right)d\lambda,
	\label{eq:image}
\end{align}
where $A(u, v)$ is the 2D Fourier transform of $a(x, y)$.
The key observation from (\ref{eq:rainbow}), (\ref{eq:image}) is that a coded aperture placed on plane P2 causes a spectral blur given by $a(x, y)$ and a spatial blur given by $\left|A\left(-\frac{x}{\lambda f}, -\frac{y}{\lambda f}\right)\right|^2$.
As shown in Fig.\ \ref{fig:space_spectrum}, a slit causes a severe spatial blur, whereas an open aperture causes large spectral blur. 
The solution is to introduce an invertible blur in both domains, which can be achieved using a coded aperture, shown in the last column.
We use the same coded aperture that was used in \cite{saragadam2018krism}, as it is designed to promote invertibility in both domains.

%
%

\paragraph{Optical setup.}
Our optical setups is in principle similar to Fig.\ \ref{fig:4f}(a).
We place a spatial light modulator on the rainbow plane (P2) and sensor on spatial plane (P3) to achieve spectral filtering.
The optimized binary code \cite{saragadam2018krism} is placed in the lens assembly  
%
Figure \ref{fig:4f}(b) shows a schematic of a practical implementation of the same optical setup.
We use a Liquid Crystal on Silicon (LCoS) display as a spatial light modulator for spectral filtering.
%
%
%
%
%

\paragraph{Effect of coded aperture.}
Given the HSI of the scene, $H(x, y, \lambda)$, the coded aperture introduces spatial and spectral blurs in the following way,
\begin{align}
\widehat{H}(x, y, \lambda) &= \left(A\left(\frac{x}{\lambda f}, \frac{y}{\lambda f}\right) \ast H(x, y, \lambda)\right) \ast a(\lambda \nu_0 f, y),
\end{align} 
i.e., all operations are now performed on a modified version of the HSI of the scene. 
%
%
%
Given a spectral profile $s_k(\lambda)$, the proposed setup directly computes filtered image,:
\begin{align}
\widehat{f}_k(x, y) = \int\limits_{\lambda} \widehat{H}(x, y, \lambda) s_k(\lambda) c(\lambda) d\lambda,
\label{eq:meas}
\end{align}
by loading $s_k(\lambda)$ on the spatial light modulator.
With the optical setup in place, we will next see how to use the programmable spectral filter to perform optical classification.

\section{Learning Discriminant Filters}
\label{section:proposed}

\begin{figure}[!tt]
	\centering
	\includegraphics[width=\columnwidth]{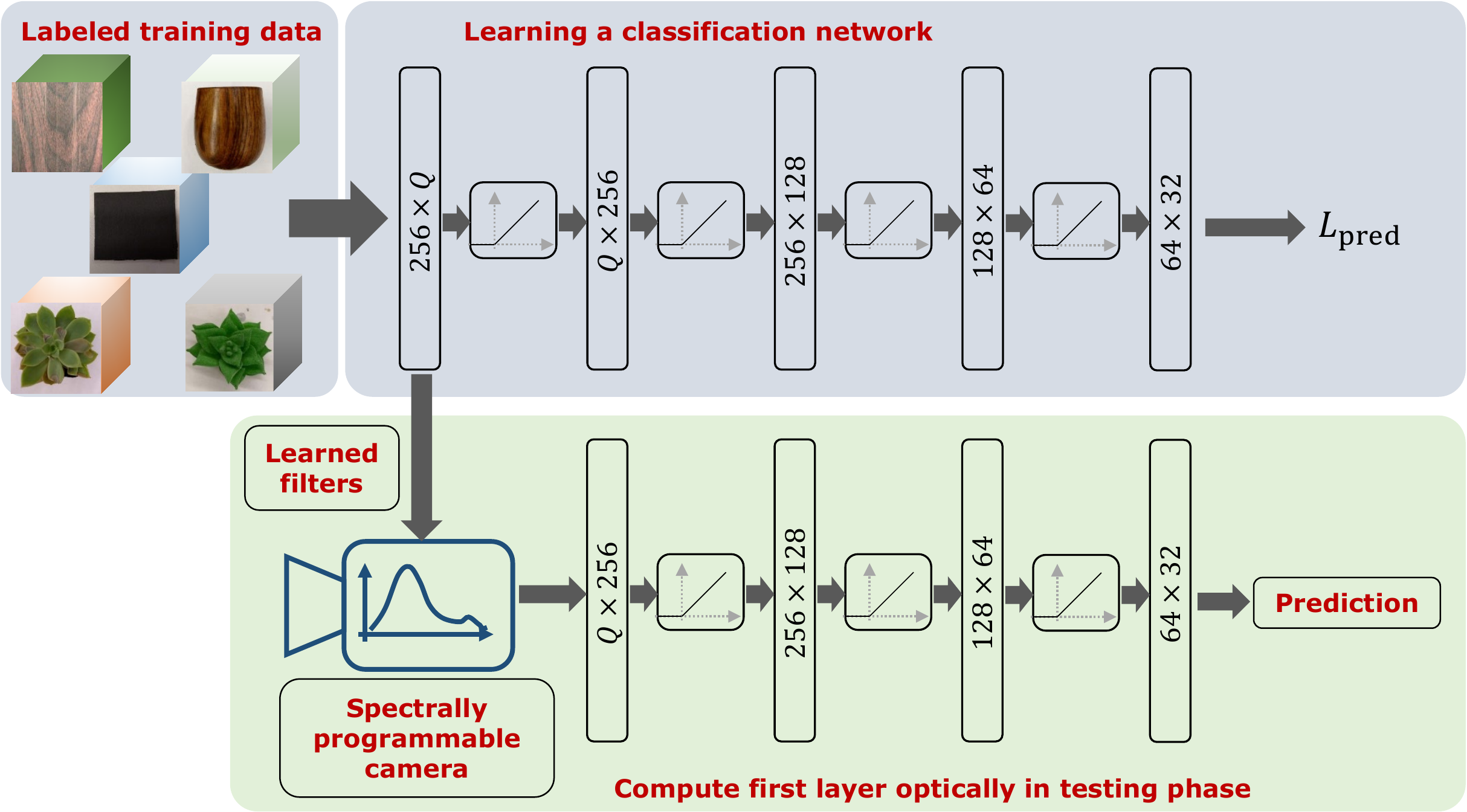}
	\caption{\small\textbf{Proposed optical classifier.} The proposed optical classifier broadly consists of two stages. In the first stage, we learn the weights of a neural network with spectrum as input and class label as output. The training process outputs the set of discerning filters, marked "learned filters" in the image. In testing stage, we filter the HSI of the scene with the learned filters, thereby replacing the first layer of the classifier with an optical implementation. This results in a high accuracy, per-pixel classifier while requiring far fewer measurements than the size of the HSI.}
	\label{fig:schematic}
\end{figure}
With camera that is capable of capturing images with arbitrary spectral profiles, we pursue two questions; one, how many filters are required for classifying $K$ classes, and two, what spectral filters maximize classification accuracy.
The questions above are closely tied to the type of classifier under consideration.
We detail the two classifier architectures we explore in this paper which help answer the questions above.
Note that any classifier which relies on the linear projection can be used. 
For the sake of exposition, we only evaluate SVM and neural networks.

\subsection{Support Vector Machine}
SVMs provide a binary, linear classifier by learning a separating hyperplane on the training dataset.
Given a set of data points $\left\{\bfx_k, y_k\right\}_{k=1}^N$, where $y_k \in \{0, 1\}$ is the label of $\bfx_k$, SVM seeks to solve the following optimization problem,
\begin{align}
	\min_{\bfw, c} \frac{1}{N} \sum_{k=1}^{k} \max(0, 1 - y_k(\bfw^\top \bfx_k + c) ) + \lambda \| \bfw \|^2,
	\label{eq:svm}
\end{align}
where $\lambda$ is a tuning parameter. The output of solving the optimization problem is the vector $\bfw$ and intercept $c$. 
In the context of optical classification, $\bfw$ is the filter that maximizes accuracy for binary decision.
For $K$-class decision, we choose a \emph{one-vs-all} classification strategy, which uses $K$ hyperplanes, and hence $K$ spectral filters.


\subsection{Deep Neural Networks}

Deep neural networks (DNNs) provide a richer alternative to SVMs.
We model the first  linear unit of the DNN to be the programmable spectral filter and train a model whose input is the spectral profile at a pixel and whose output is the material class label as a one-hot vector.
While there are many possible architectures, we choose a simple, five-layer neural network as an example with all layers being fully connected. 
Figure \ref{fig:schematic} gives a brief overview of the proposed training and testing methodology.
The weights of first fully connected layer, $A_1$ are the learned discriminating filters, and hence the first layer can be evaluated optically, thereby circumventing the need to measure the full spectrum at each pixel.
%
%
The number of filters, $Q$ depends on the number of materials and how easily they can be separated. In our experiments, we classified a total of $5$ objects.
We then varied the number of filters and computed mean classification accuracy. Based on this, we picked the optimal number of filters.
%
%
We note that the idea of optically computing the first layer has been explored before in the context of designing color filter arrays \cite{chakrabarti2016learning} and processing light fields  \cite{chen2016asp}.

\subsection{Simulations.}
\begin{figure}[!tt]
	\centering
	\includegraphics[width=\columnwidth]{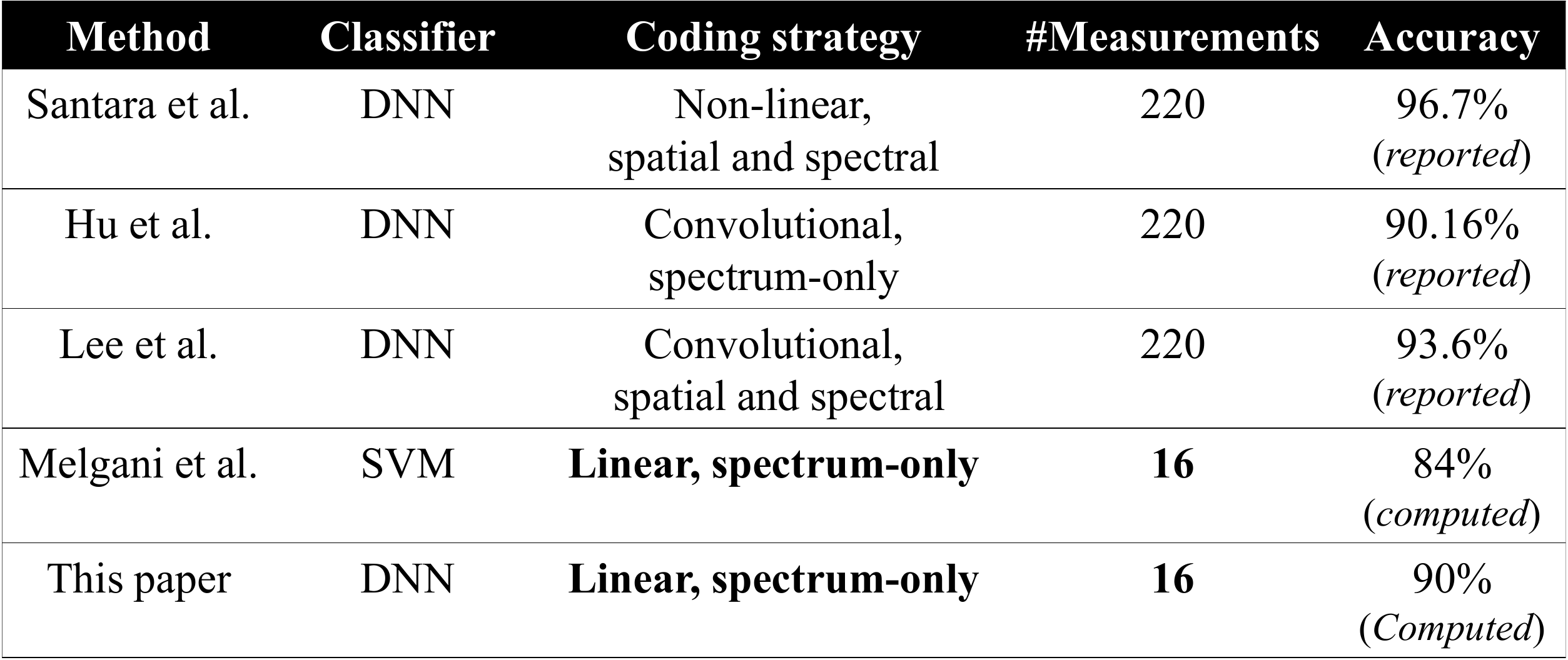}
	\caption{\small\textbf{Simulations on the Indian Pines dataset.} We compare state-of-the-art classifiers against the classifiers proposed in this paper. By \emph{reported} we report the accuracy figures listed in the respective papers, while \emph{computed} results were generated by us. A key feature of our optical setup is that it can only compute linear projections of spectra. While this leads to reduction in accuracy, the number of captured images are far fewer.}
	\label{fig:indian_pines_sim}
\end{figure}

We compare SVM and the 5-layer DNN classifier to some of the state-of-the-art techniques in spectral-classification on the NASA Indian Pine dataset which consists of $220$ spectral bands with $16$ object classes.
%
%
Figure \ref{fig:indian_pines_sim} tabulates the accuracies with classifiers used in this paper in bold.
We observe that the accuracy is lower than state-of-the-art, which is expected as we only use spectral information, while the other techniques use both spatial and spectral information.
However, relying on a spectrum-only classifier lets us capture far fewer images than the number of spectral bands.

\section{Experiments}
\label{section:real}
%
%
We demonstrate capabilities of our setup  for video-rate binary classification with binary SVM as well as matched filtering, and multi-class classification with  multi-class SVM and DNNs.
\begin{figure}[!tt]
	\includegraphics[width=\columnwidth]{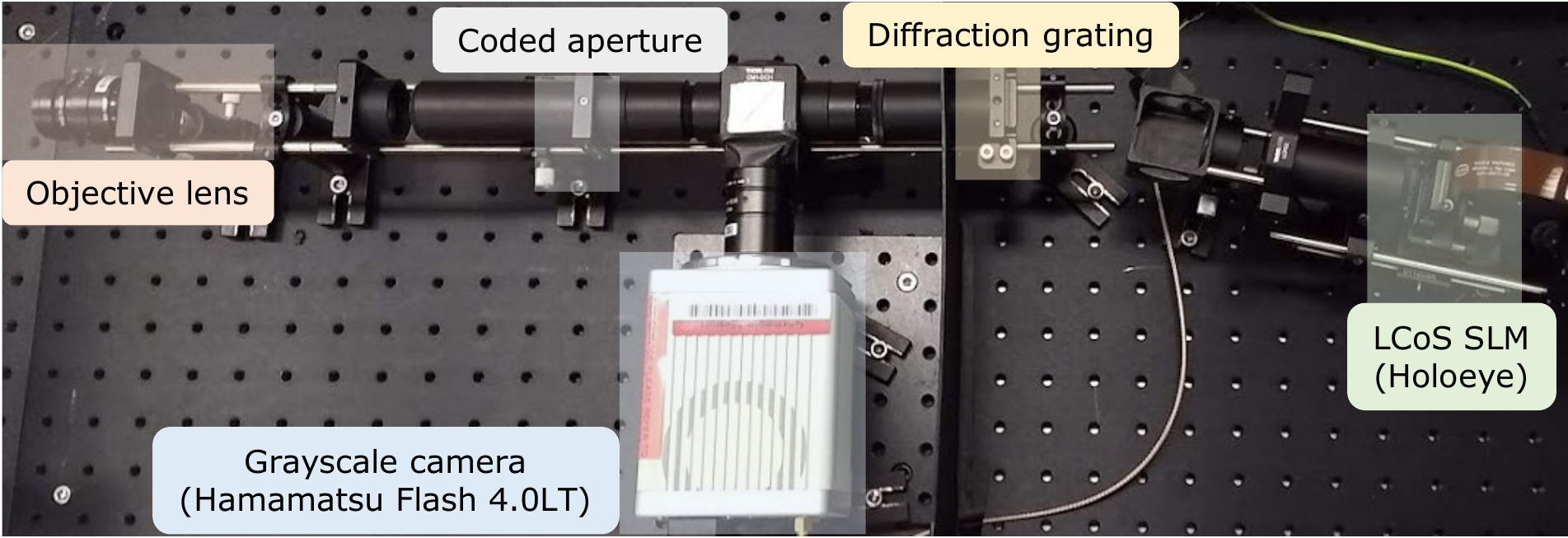}
	\caption{\textbf{Lab prototype.} The picture shows the lab prototype we built with only the major components marked.  We used an objective lens of $8$mm focal length, while all other lenses were $100$mm.}
	\label{fig:prototype}
\end{figure}

\begin{figure}[!tt]
	\centering
	\begin{subfigure}[c]{0.22\columnwidth}
		\centering
		\includegraphics[width=\textwidth]{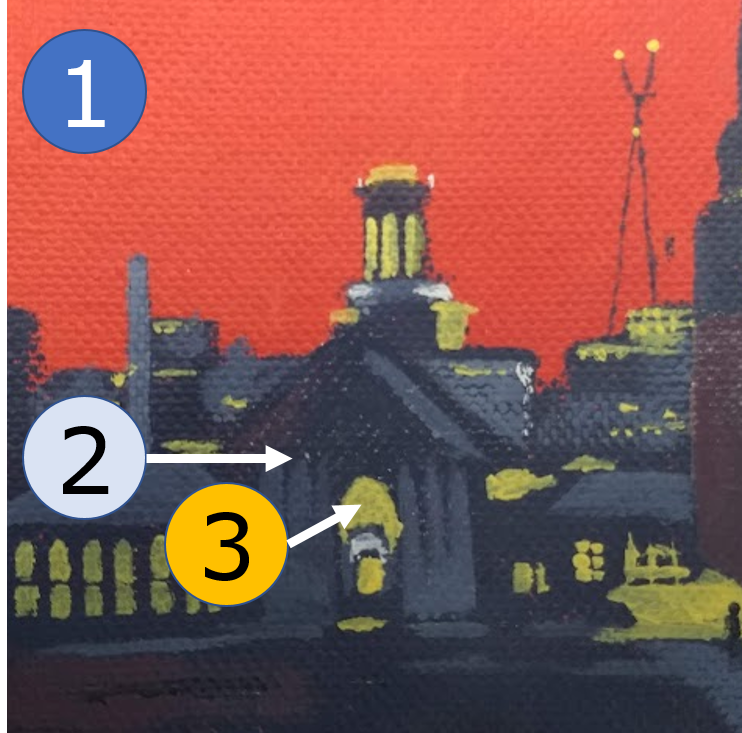}
		\caption{RGB}
	\end{subfigure}
	\hspace{0.05em}
	\begin{subfigure}[c]{0.22\columnwidth}
		\centering
		\includegraphics[width=\textwidth]{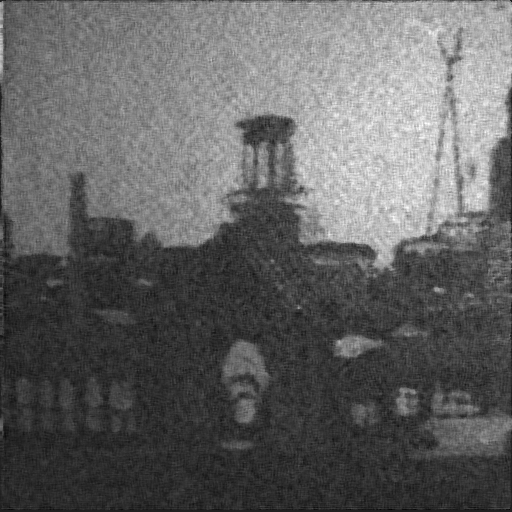}
		\caption{$650$nm}
	\end{subfigure}
	\hspace{0.05em}
	\begin{subfigure}[c]{0.22\columnwidth}
		\centering
		\includegraphics[width=\textwidth]{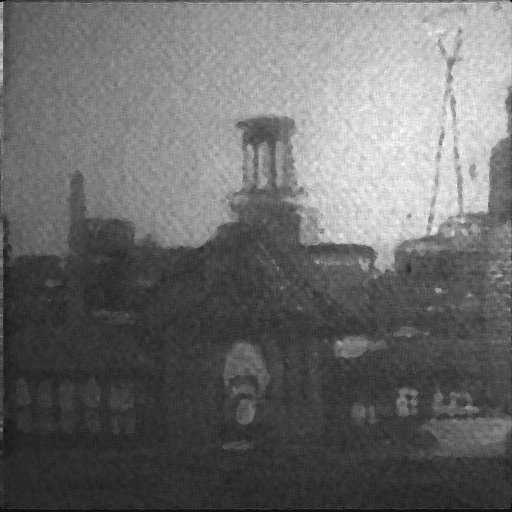}
		\caption{$750$nm}
	\end{subfigure}
	\hspace{0.05em}
	\begin{subfigure}[c]{0.22\columnwidth}
		\centering
		\includegraphics[width=\textwidth]{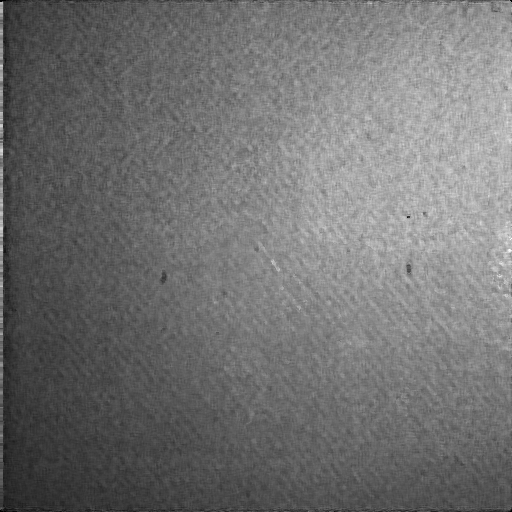}
		\caption{$850$nm}
	\end{subfigure}
	\\
	\begin{subfigure}[c]{\columnwidth}
		\centering
		\includegraphics[width=\textwidth]{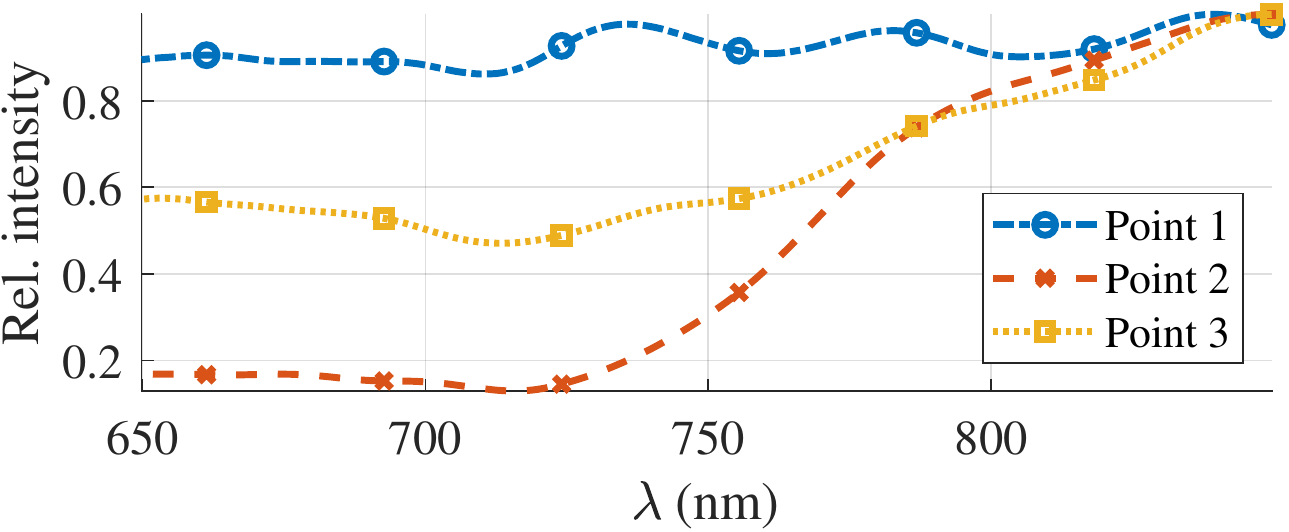}
	\end{subfigure}
	\caption{\textbf{Example HSI.} Our prototype is designed to capture images from $600$nm to $900$nm. (a) was captured using a cellphone while (b)-(d) are images captured by our setup. Bottom row shows spectral profiles at three marked points. Note how all the pigments disappear at $\lambda = 850$nm in (d).}
	\label{fig:example}
\end{figure}
\begin{figure}[!tt]
	\centering
	\begin{subfigure}[c]{\columnwidth}
		\includegraphics[width=\columnwidth]{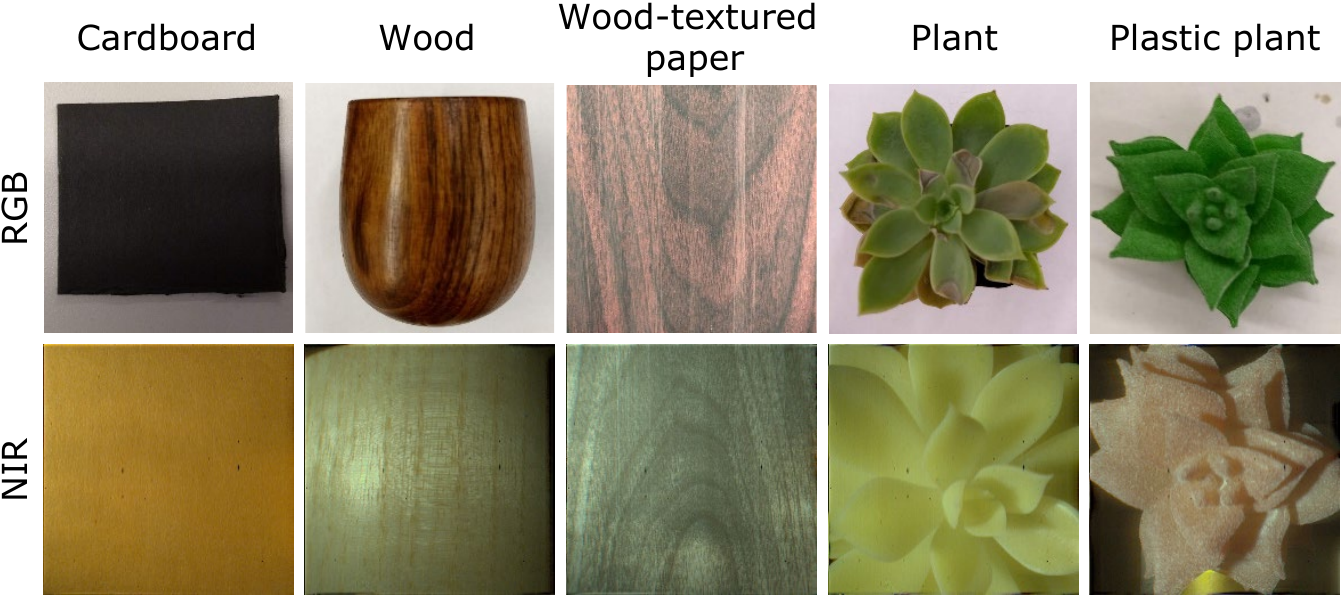}
		\caption{Visible and NIR images.}
	\end{subfigure}
	\\
	\begin{subfigure}[c]{\columnwidth}
		\includegraphics[width=\columnwidth]{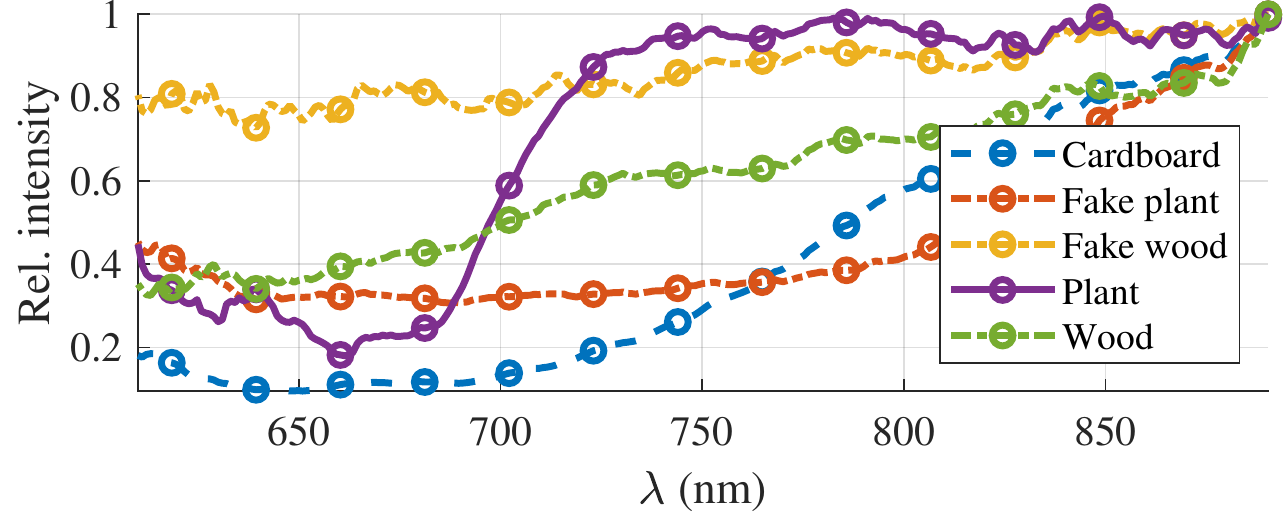}
	\end{subfigure}	
	\caption{\textbf{Material dataset.} The figure shows false colored images of the 5 different materials we collected for classification purpose. (b) shows average spectra of the materials as measured by our lab prototype.}
	\label{fig:materials}
\end{figure}
\paragraph{Optical setup.}
Figure \ref{fig:prototype} shows a photograph of the lab prototype we built along with labels for relevant components.
A detailed optical layout along with the list of components is in appendix.
%
 %
Our SLM is a Holoeye LCoS SLM with a frame rate of 60 Hz that works as a secondary monitor.
We used an NIR-sensitive sCMOS camera (Hamatasu ORCA Flash 4.0 LT).
In order to classify materials accurately, we designed our system to image from $600$nm to $900$nm, which is the near infrared (NIR) regime.
Our setup is capable of coding spectrum at a resolution of $3.3$nm, giving us $100$ spectral bands.
Finally, the SLM acts as a dynamic spectrally-selective camera and hence can be directly used for measuring the complete HSI. 
To do so, we display permuted Hadamard patterns on the SLM to capture a $512\times512\times256$ dimensional HSI.
Figure \ref{fig:example} shows an example of captured HSI of an acrylic painting.

\begin{figure}[!tt]
	\centering
	\begin{subfigure}[c]{0.48\columnwidth}
		\centering
		\includegraphics[width=\columnwidth]{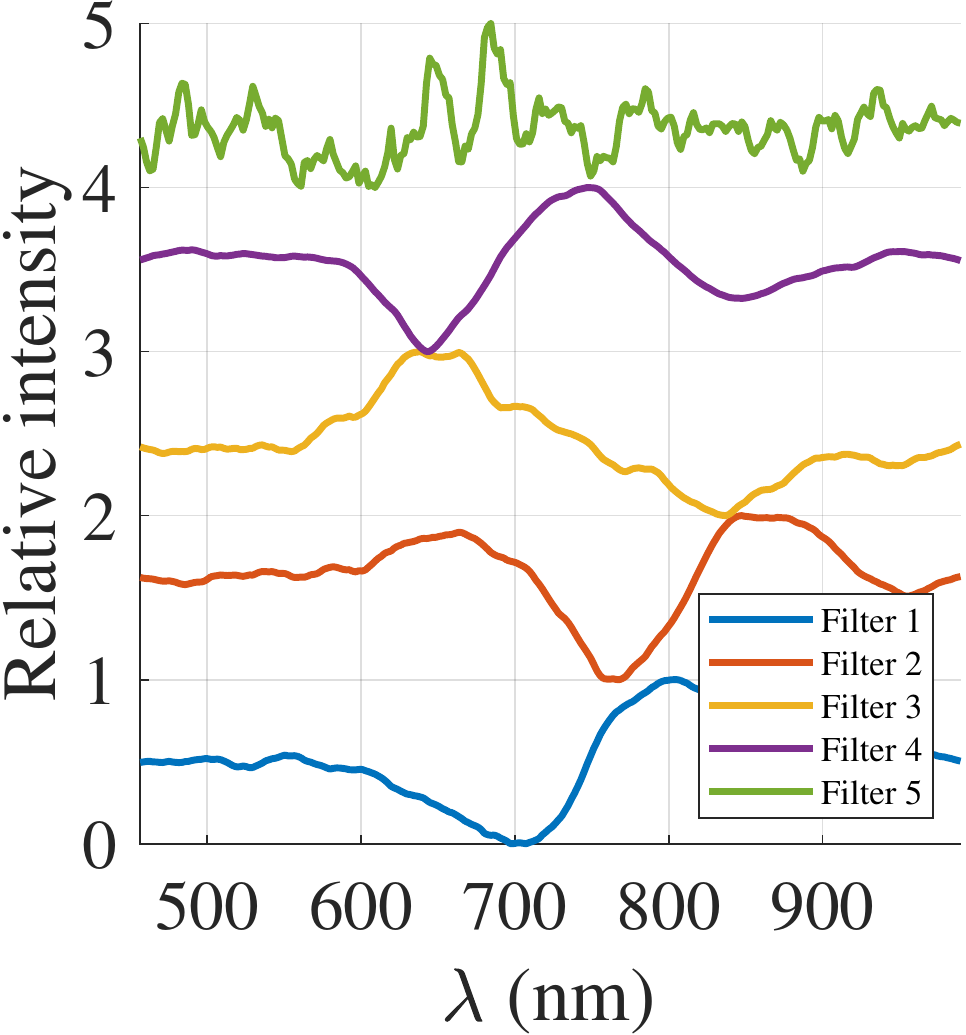}
		\caption{SVM filters}
	\end{subfigure}
	\begin{subfigure}[c]{0.48\columnwidth}
		\centering
		\includegraphics[width=\columnwidth]{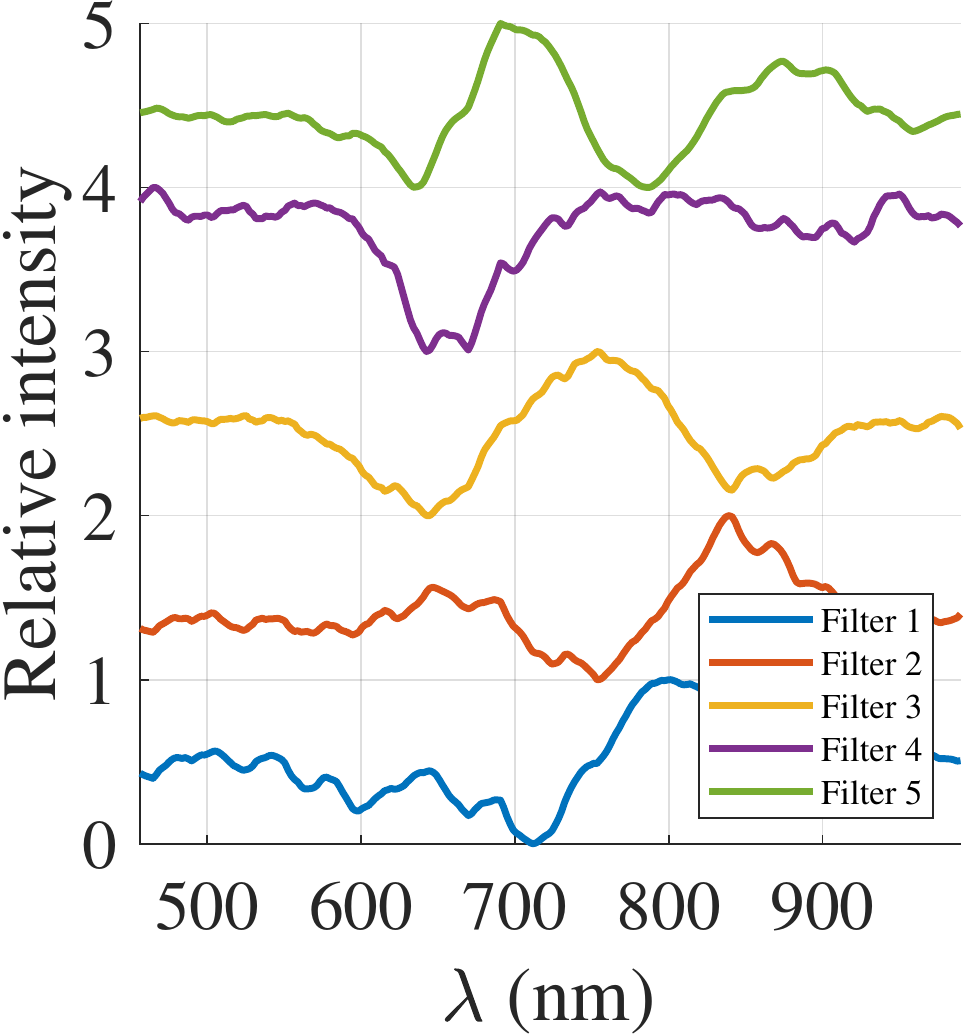}
		\caption{DNN filters}
	\end{subfigure}
	\caption{\textbf{Learned filters.} The output of a multi-class SVM is $K$ separating hyperplanes, which results in $K$ filters, shown in (a). Similarly, the DNN architecture consists of several layers, of which the first layer is linear. Hence the training process results in weights shown in (b) that can be used as spectral filters.}
	\label{fig:filters}
\end{figure}

\begin{figure}[!tt]
	\centering
	\begin{subfigure}[t]{0.3\columnwidth}
		\centering
		\includegraphics[width=\textwidth]{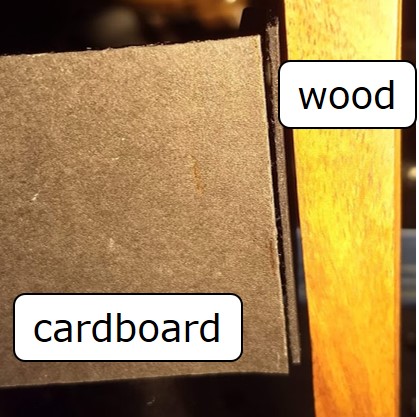}
		\caption{RGB image}
	\end{subfigure}
	\hspace{0.01em}
	\begin{subfigure}[t]{0.3\columnwidth}
		\centering
		\includegraphics[width=\textwidth]{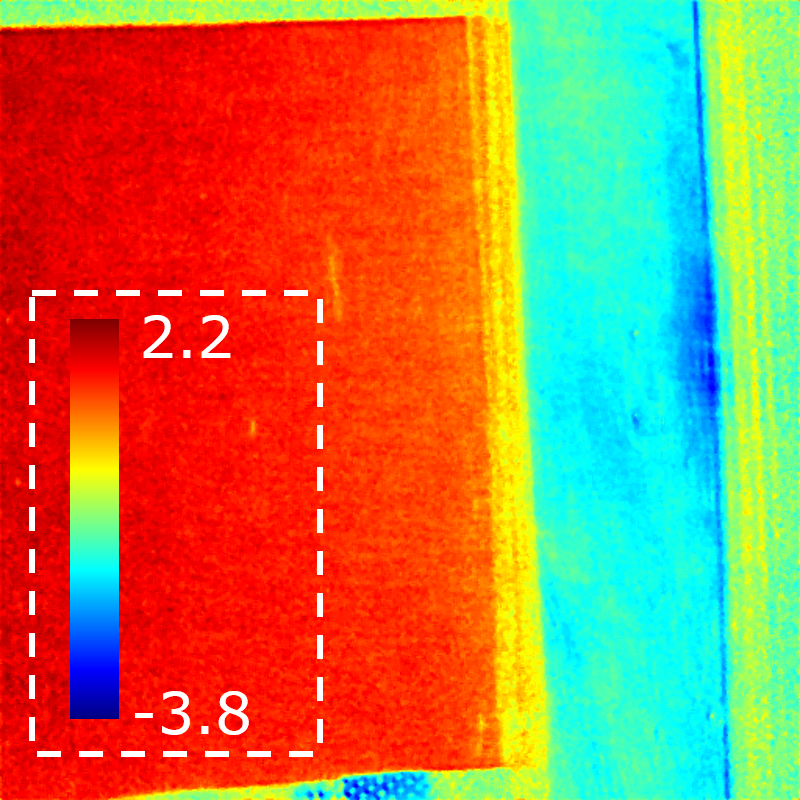}
		\caption{Full HSI scan + projection\\(256 meas.)}
	\end{subfigure}
	\hspace{0.01em}
	\begin{subfigure}[t]{0.3\columnwidth}
		\centering
		\includegraphics[width=\textwidth]{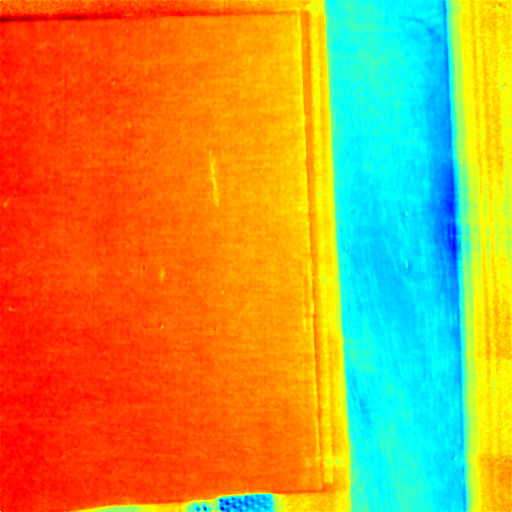}
		\caption{optical projection\\(2 meas.)}
	\end{subfigure}
	\\
	\begin{subfigure}[t]{\columnwidth}
		\centering
		\includegraphics[width=\textwidth]{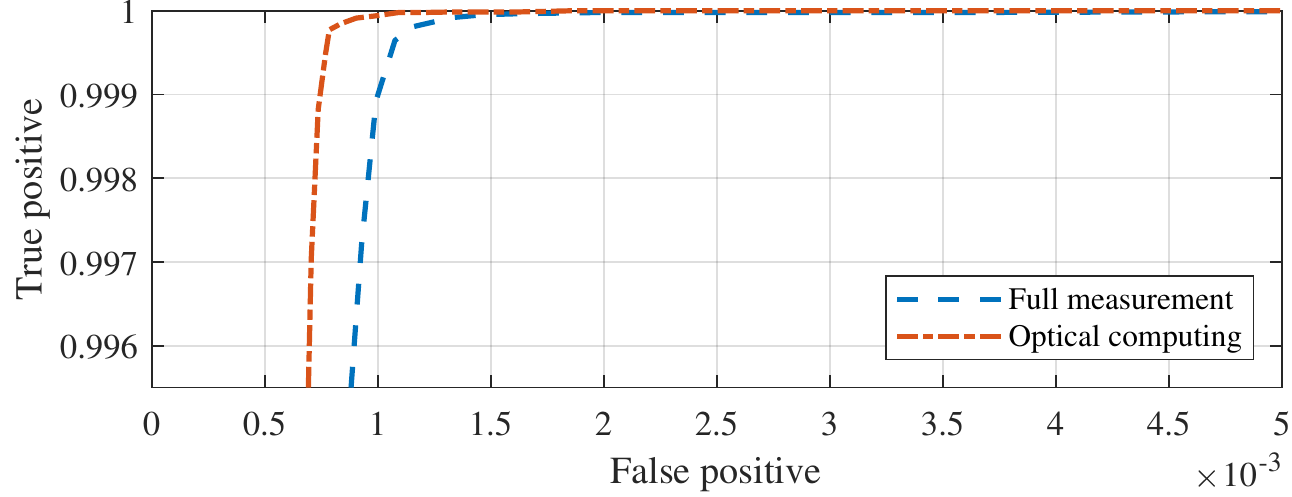}
	\end{subfigure}
	\caption{\textbf{Advantage of optical computing.} We show an example of binary classification between cardboard and wood (a) using per-pixel SVM. Optical computing achieves higher accuracy with far fewer measurements.}
	\label{fig:roc}
\end{figure}

\vspace{-5mm}
\paragraph{Calibration.}
Our optical setup broadly requires calibration of the code resulting in spectral blur, calibration of wavelengths and finally, spatial PSF. 
We use narrow-band lasers for  calibrating both code and wavelengths, and use a $10\mu$m pinhole for calibrating spatial PSF. 
Details  are available in the appendix.
%
\begin{figure}[!tt]
	\centering
	\includegraphics[width=\columnwidth]{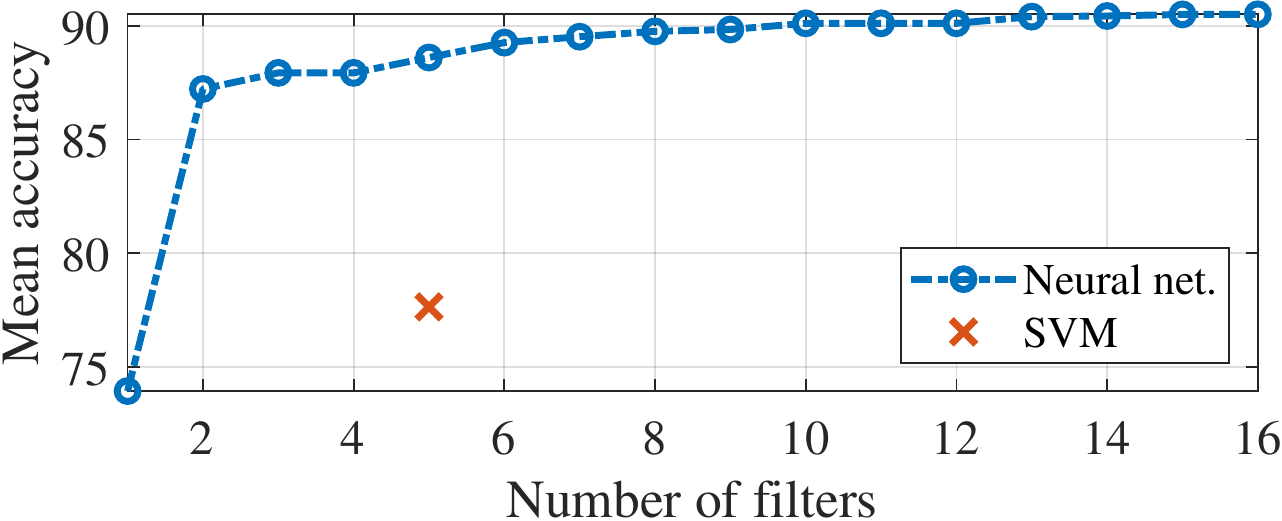}
	\caption{\textbf{Accuracy vs. number of filters.} The plot shows accuracy as a function of number of filters. The accuracy increases initially and then saturates. We hence use the knee point of the curve as the optimal number of filters.} 
	\label{fig:acc_vs_nfilters}
\end{figure}
\begin{figure}[!tt]
	\centering
	\begin{subfigure}[t]{0.31\columnwidth}
		\centering
		\includegraphics[width=\textwidth]{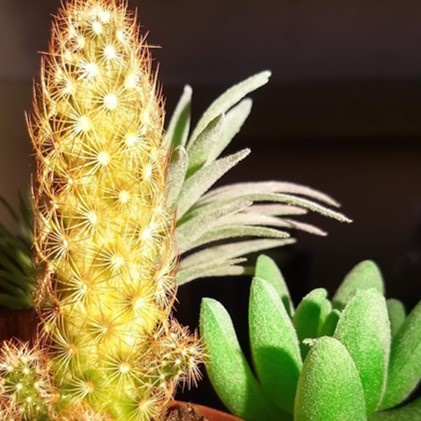}
		\caption{RGB}
	\end{subfigure}
	\hspace{0.05em}
	\begin{subfigure}[t]{0.31\columnwidth}
		\centering
		\includegraphics[width=\textwidth]{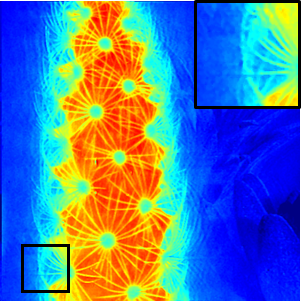}
		\caption{SVM score}
	\end{subfigure}
	\hspace{0.05em}
	\begin{subfigure}[t]{0.31\columnwidth}
		\centering
		\includegraphics[width=\textwidth]{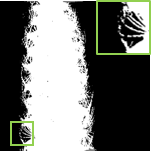}
		\caption{Label}
	\end{subfigure}
	\caption{\textbf{Per-pixel classification.} Due to per-pixel operation with high spatial resolution, our imager can clearly identify the micro-structures such as the cactus thorns by capturing only two images instead of the complete HSI.}
	\label{fig:cactus}
\end{figure}

\begin{figure*}[!tt]
	\centering
	\includegraphics[width=\textwidth]{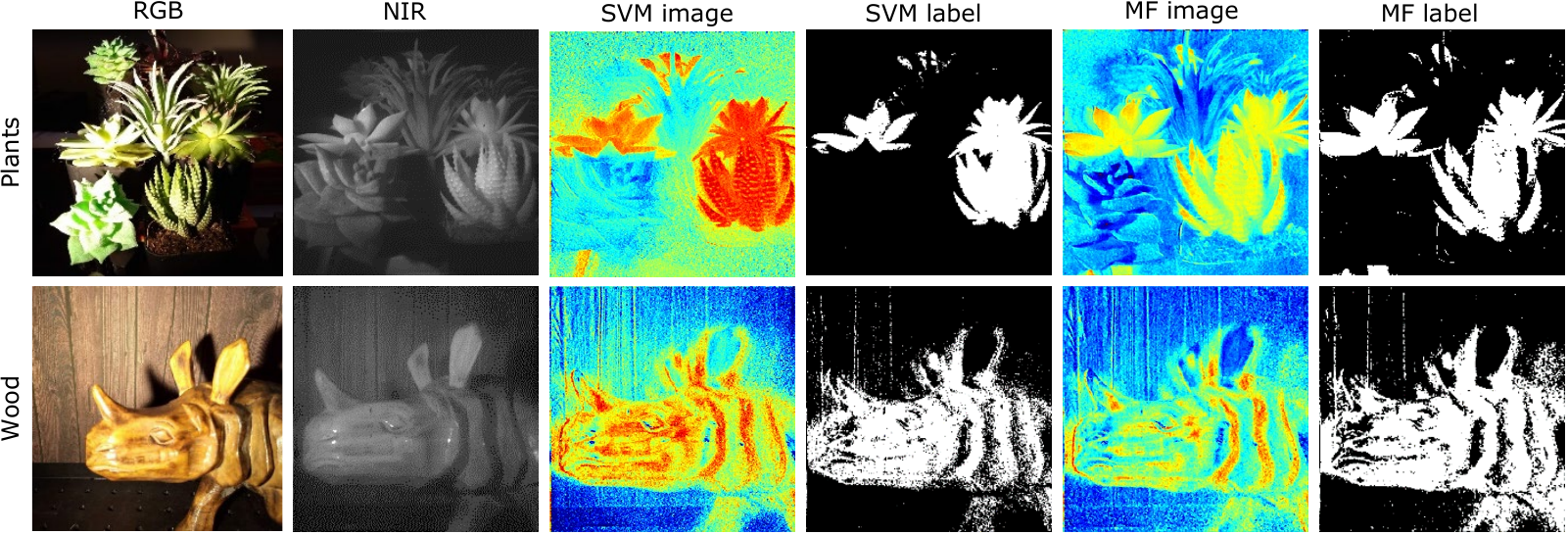}
	\caption{\textbf{Various binary classifiers.} We compare binary classification using SVM and matched filtering (MF). First row is a comparison of real wood (rhino) and fake wood (background, printed paper), while the second row is real and fake plants. Due to dynamic programming capability, we can classify with arbitrary filters and hence utilize any classifier that relies on linear projection of spectrum at each pixel.}
	\label{fig:binary_svm}
\end{figure*}
%
\paragraph{Dataset.}
We show classification results with a total of five types of subjects: 1) black cardboard, 2) varnished wood, 3) wood-textured paper, 4) real plants, and 5) plastic plants.
The choice of objects stems from similarity of these materials (plants vs plastic plants) in visible wavelengths, while having distinctly different spectra in NIR domain.
We collect one HSI for each of the materials and manually label them, giving a total of 5 HSI for training.
Figure \ref{fig:materials} shows visible and false-NIR images, as well as the average spectrum for each material.
We note that none of the objects used for training the classifiers were reused in testing phase.
\begin{figure*}[!tt]
	\centering
	\includegraphics[width=\textwidth]{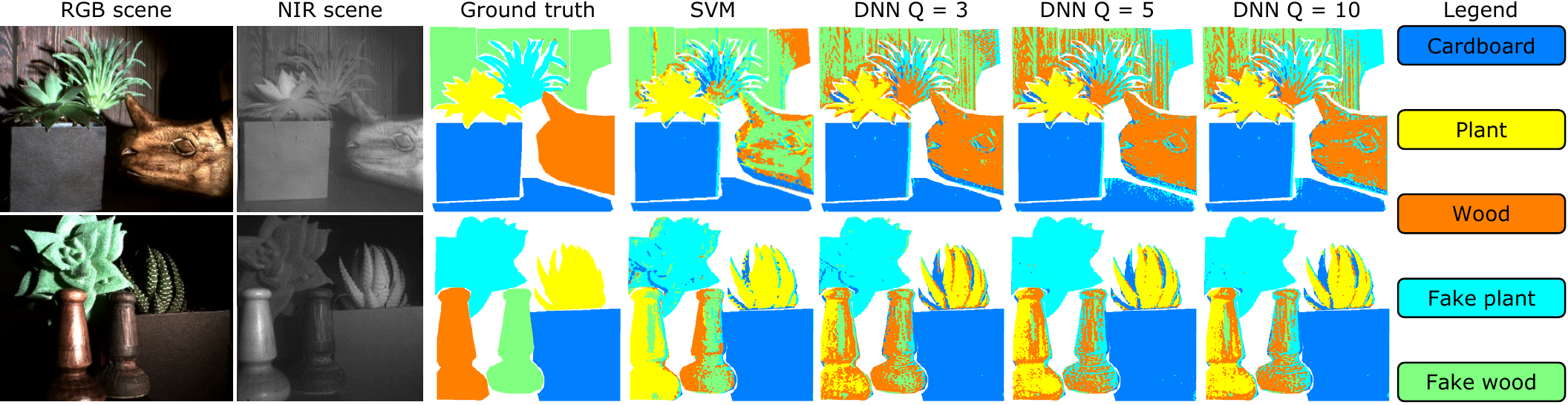}
	\caption{\textbf{Optical classification.} We show two examples of classification where the linear operations are directly computed in the optical domain. The ground truth labels were obtained by hand annotation. SVM required a total of 11 measurements for five filters, whereas DNN with 3, 5, 10 filters required 7, 11, 21 images respectively. The RGB images shown how the objects are not easily discernable in the visible domain, while they are accurately identified in the NIR domain along with optical classification. }
	\label{fig:optical_classification}
\end{figure*}

\vspace{-5mm}
\paragraph{Training classifiers.}
We trained two classifiers -- multi-class SVM and DNN with varying number of filters.
For SVM, we used Scikit-Learn \cite{scikit-learn} in a one-vs-all configuration which learned a total of $5$ spectral filters.
The learned spectral filters are shown in Fig.\ \ref{fig:filters} (a)

DNNs were trained with the network architecture shown in Fig.\ \ref{fig:schematic} with loss function set to cross entropy. The number of spectral filters were varied from $1$ to $20$ to compare performance.
We learned the network using the PyTorch framework \cite{paszke2017automatic} with learning rate set to $10^{-3}$ for a total of $50$ epochs.
We then extracted weights of first layer and used them as spectral filters.
The learned filters are shown in Fig.\ \ref{fig:filters} (b).
Further details about the learning process are included in appendix.

Figure \ref{fig:acc_vs_nfilters} shows a plot of accuracy as a function of number of filters, $Q$.
Accuracy of the classifier increases sharply initially and then saturates which implies that more spectral filters do increase accuracy but there is diminishing returns after a point.
Based on this, we used $3, 5, 10$ filters for comparisons in our real experiments.
%


\paragraph{Handling scale of features.}
A key requirement of any classifier is that the scale of features be same during training and testing.
A common practice is to set the norm of feature at $(x_0, y_0)$,  $\|H(x_0, y_0, \lambda)\|$ to unity, or the maximum value to unity. 
In our case, this requires having knowledge of the complete spectral profile, which defeats the purpose of optical computing.
instead, we normalize our measurements with sum of the spectrum, $\int_\lambda H(x_0, y_0, \lambda)$, which can be measured by displaying a spectral profile with all ones.
The measured featured vectors are then,
\begin{align}
	I_\text{sum}(x_0, y_0) &= \int_\lambda H(x_0, y_0, \lambda)d\lambda\\
	\widetilde{I}_k(x_0, y_0) &= \int_\lambda H(x_0, y_0, \lambda)s_k(\lambda) d\lambda\\
	I_k (x_0, y_0) &= \frac{\widetilde{I}_k(x_0, y_0)}{I_\text{sum}(x_0, y_0)}
\end{align}
We scale the spectra the same way even while training, which makes the scaling consistent.
Hence any set of measurements with spectral profiles requires one extra image.

\begin{figure}[!tt]
	\centering
	\begin{subfigure}[c]{0.48\columnwidth}
		\centering
		\includegraphics[width=\textwidth]{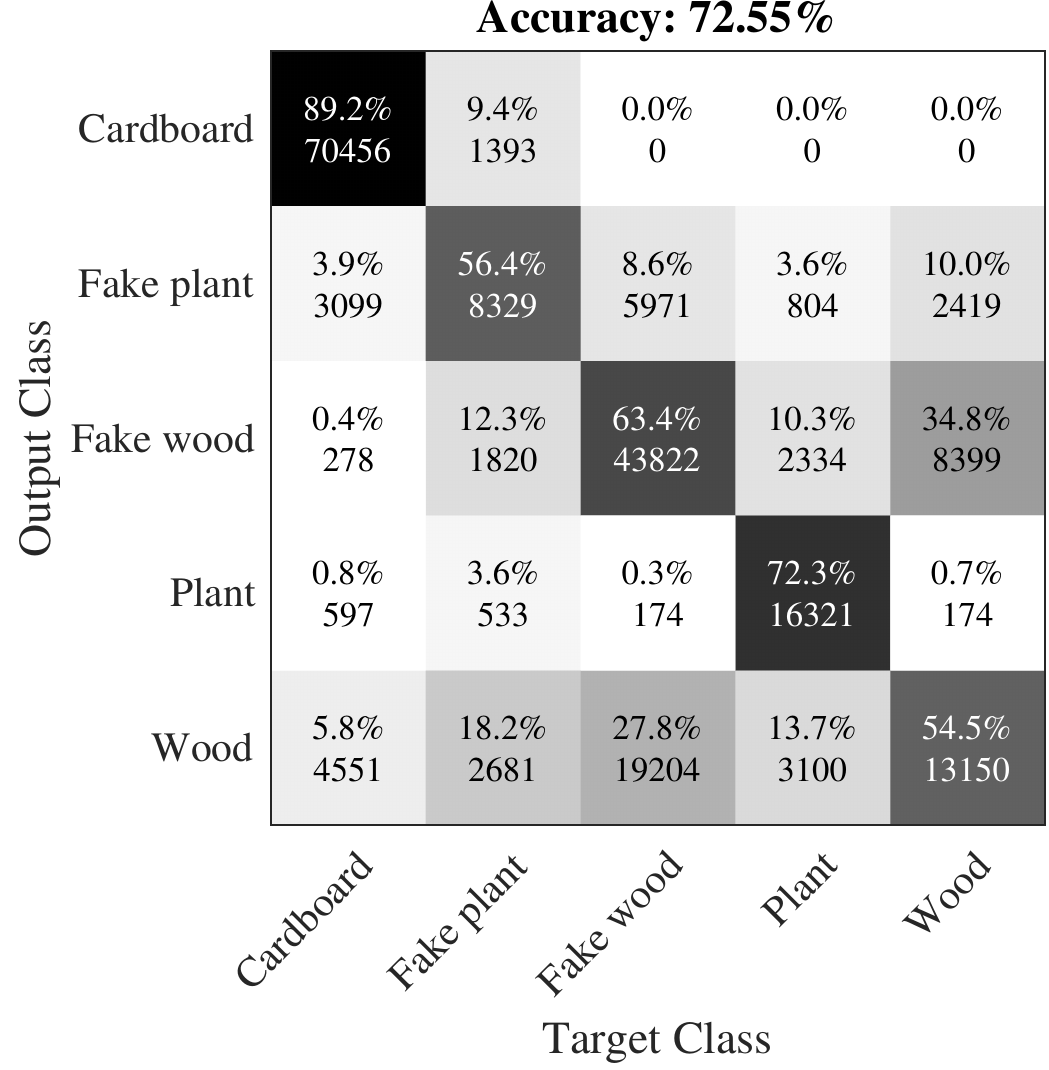}
		\caption{SVM}
	\end{subfigure}
	\begin{subfigure}[c]{0.48\columnwidth}
		\centering
		\includegraphics[width=\textwidth]{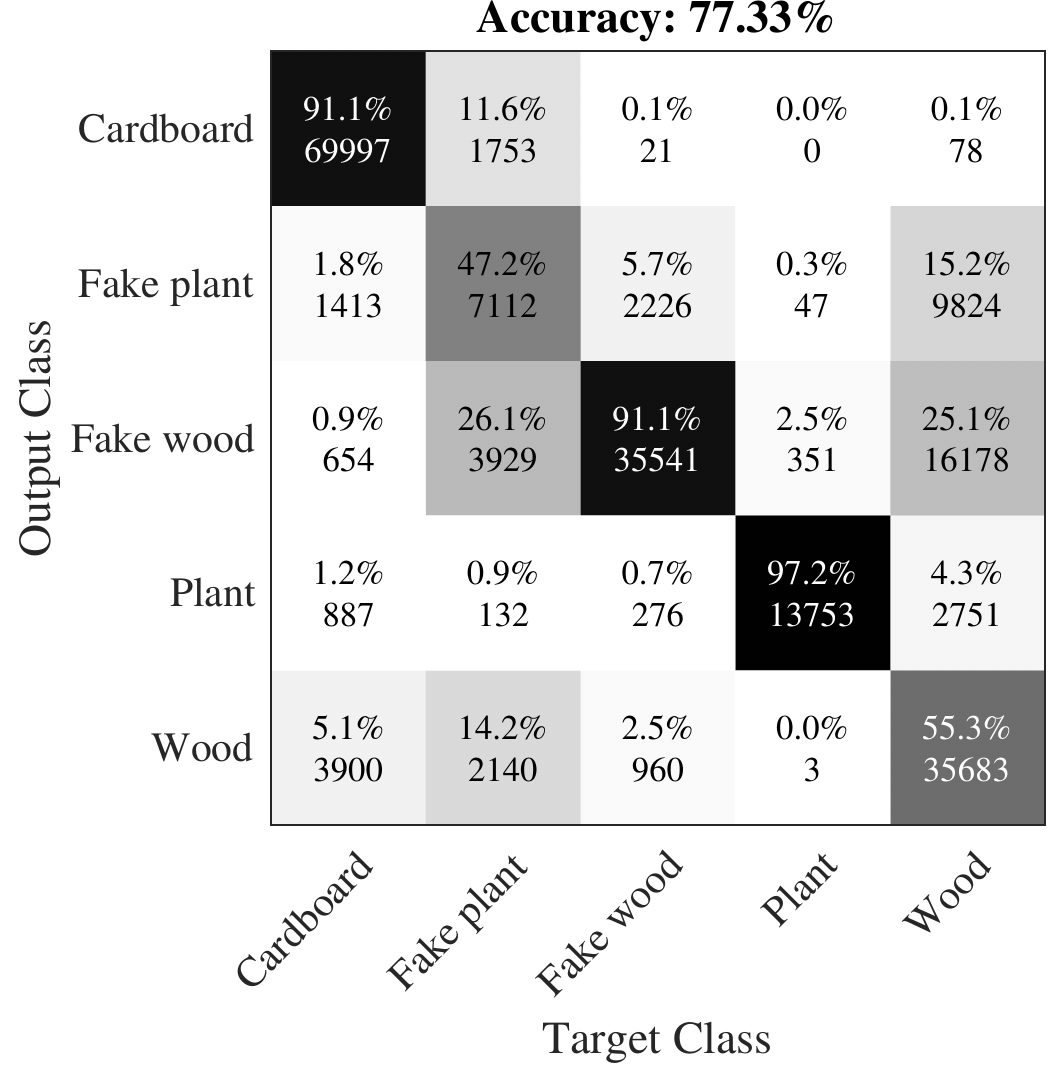}
		\caption{Neural net.}
	\end{subfigure}
	\caption{\textbf{Confusion matrices for classifiers.} Neural networks typically outperform SVM. Among object classes, ``wood" and ``fake wood" get confused the most, as their spectra are similar. In contrast, ``cardboard" is most different from all other spectra and hence has high accuracy.}
	\label{fig:confusion_mat}
\end{figure}
\paragraph{Binary classification.}
The simplest task possible with our optical setup is a binary classification, where the label at each pixel belongs to one of the two possible classes. 
In such a situation, one may either use a linear SVM where the spectral filter is the learned supporting hyperplane, $\bfw$, or use a matched filter, where the spectral filter is difference of spectra of the two classes, $s_1(\lambda) - s_2(\lambda)$.
Figure \ref{fig:roc} evaluates the advantages of optical classification. (b) visualizes the SVM score at each pixel obtained by scanning the complete HSI and then computing the projection to the SVM hyperplane, which requires a total of 256 measurements. In contrast, optical projection, shown in (c) requires only two images. Bottom row shows the Receiver operating Characteristic (RoC) of classification for both cases. The SNR advantage is evident; the area under the curve for optical projection ($0.7194$) is higher than full measurement and then projection ($0.7912$).
Figure \ref{fig:cactus} shows classification of a real cactus surrounded by several plastic plants. The SVM score in (b) as well as the labels show that our setup is capable of resolving very thin structures such as the cactus thorns.
Figure \ref{fig:binary_svm} shows classification results for real vs. plastic plants and real vs. fake wood with binary SVM as well as matched filtering.
Note that the objects are not easily discernable in RGB domain, while they are easily isolated after spectral filtering.
%
Figure \ref{fig:teaser} shows a video rate classification of a real plant and a fake plant.
The video was captured at $4$ frames per second with alternating spectral profile and all ones pattern.
Note how the real plant is tracked across all frames, while the fake plant is ignored.

\paragraph{Multi-class classification.}
We test the SVM and DNN filters learned on training data to classify a scene made of various materials from the set of five materials.
Figure \ref{fig:optical_classification} shows classification results for two scenes for various techniques.
The ground truth annotation was obtained by manually annotating the objects, and then this was used for measuring the accuracy of classification.
Visually, DNNs outperform SVM, as is visible from the accurate classification of the wooden rhinoceros head.
Figure \ref{fig:confusion_mat} shows a confusion matrix for SVM and DNN with 5 filters.
The accuracies are not very high as we depend on spectral features alone. Accuracy can be significantly increased if spatial information is used along with spectral profiles.
This is done by first capturing the $Q$ images and then using the spatial information to classify. 


\section{Discussions and Conclusion}
\label{section:discussions}
We propose a per-pixel material classifier that relies on a high resolution programmable spectral filter.
We achieve this by learning spectral filters that can achieve high classification accuracy and then measure images of the scene with the learned filters.
Owing to a simple, per-pixel decoding strategy, we can achieve classification at video rates.
We showed several compelling real world examples with emphasis on binary video-rate and multi-class classification.

\paragraph{Limitations.}
A key limitation of our setup is the assumption that the pixels come from a single material class.
Some real world examples are made of a mixture of materials at each class, an example being land cover.
In such a case, outputting just a class label may not suffice but relative probabilities of each class is desired.
This can be achieved by modifying the classifiers to output a score for each material at each pixel instead of most probable class.
%


\section{Acknowledgment}
\label{section:ack}
The authors acknowledge support from the National Science Foundation under the CAREER award CCF-1652569, the Expeditions award IIS-1730147, and the National Geospatial-Intelligence Agency’s Academic Research Program (Award No. HM0476-17-1-2000).

\begin{appendices}
	
\section{Theoretical background}\label{section:theory}
\subsection{Image formation model.}
We specified a simplified version of image formation model where we said that the HSI of the scene can be represented as $H(x, y, \lambda)$.
We discuss a more precise model here.

Consider the scene's spectral reflectance function, $H_R(x, y, \lambda)$, where we assume that each point in 3D space is well modeled by Lambertian reflectance.
Let $L(\lambda)$ be the spectral distribution of a spatially uniform light source.
The HSI of the scene under this illumination is then given by,
\begin{align}
	H_o(x, y, \lambda) &= H_R(x, y, \lambda) L(\lambda),
\end{align}
which was the signal model we used in the main paper. 
Then, given a camera with spectral response $C(\lambda)$, the measured image is,
\begin{align}
	I(x, y) &= \int_\lambda H_o(x, y, \lambda) C(\lambda) d\lambda \nonumber\\
			&= \int_\lambda H_R(x, y, \lambda) L(\lambda) C(\lambda) d\lambda
\end{align}
From the above equation, we see that the camera measures spectral albedo of the scene's HSI and not the spectral reflectance of models.
However, this is not a problem, as long as the light's spectral distribution is known \emph{a priori}.



\section{Learning details}\label{section:learning}
We provide details about our training process with emphasis on choice of parameters and hyperparameters. We captured a total of $1,000,000$ spectral profiles over $5$ material types. For each classifier, we used $20\%$ for training, $5\%$ for validation and $80\%$ for testing. 
We found that the testing accuracy did not improve even if we used more than $20\%$ data.

\paragraph{Support Vector Machine.}
We used the function \verb|LinearSVC| from Scikit-Learn \cite{scikit-learn} for training a \emph{one-vs-all} SVM. The only hyperparamter of tuning was penalty for the hyperplanes, $C$, which was tuned by performing a grid search over the log space from $10^{-5}$ to $1$. Hyperparamter search was done through a 3-fold cross-validation. 

\begin{table}[!tt]
	\centering
	\includegraphics[width=\columnwidth]{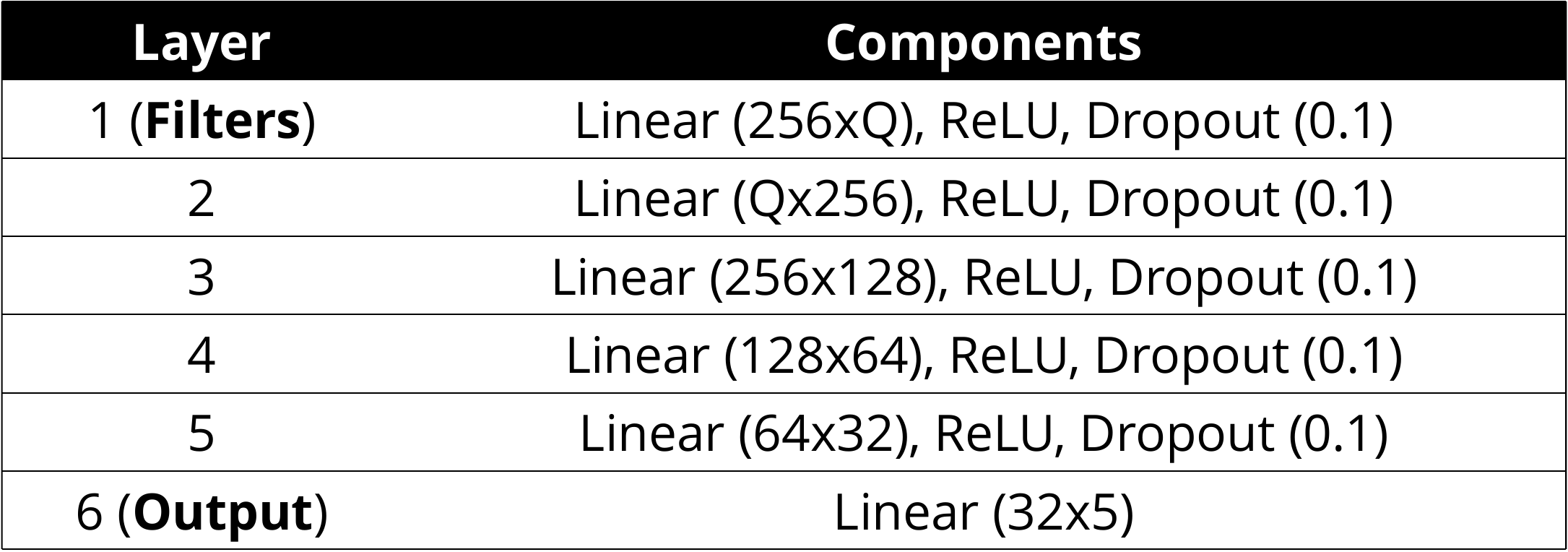}
	\caption{\small\textbf{Components of our DNN classifier.} All the layers are formed of fully connected layers with a ReLU and dropout added after each linear layer. Here, $Q$ is the number of spectral filters and was variable in our experiments to compare performance. The output was a single linear layer. During training process, we used cross-entropy as loss function.}
	\label{tab:nn_components}
\end{table}

\paragraph{Neural Networks.}
We used PyTorch \cite{paszke2017automatic} for training our neural network (DNN) classifiers. The architecture used for learning is shown in \ref{fig:schematic} and the details of each layer is provided in Table \ref{tab:nn_components}.
$Q$ is the number of filters and was varied from $1$ to $20$ to evaluate performance as a function of measurements.
We trained the network with an initial learning rate of $10^{-3}$ and trained for a total of $60$ epochs.
The filters were initialized with a principal component analysis (PCA) decomposition of training data. This lead to smoother filters and higher accuracy.
For each $Q$, we picked the model with best accuracy on validation dataset.


\section{Hardware details}\label{section:calibration}

\begin{figure*}[!tt]
	\includegraphics[width=\textwidth]{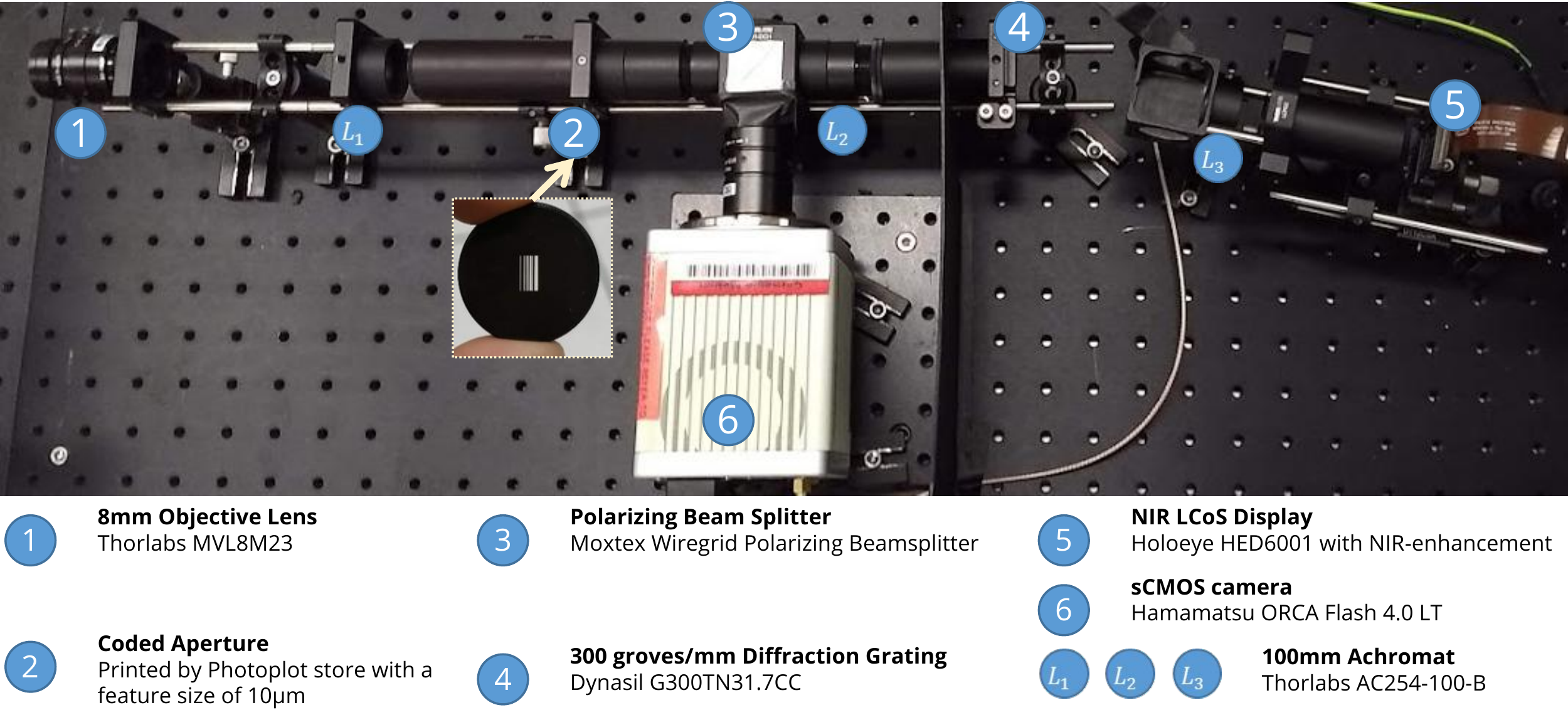}
	\caption{\textbf{Lab prototype.} A picture of the lab prototype along with major components marked with details. We skipped details about opto-mechanical components such as cage plates and posts to avoid clutter. The inset image shows the printed mask we used as coded aperture.}
	\label{fig:full_setup}
\end{figure*}

\begin{figure}[!tt]
	\centering
	\begin{subfigure}[t]{0.45\columnwidth}
		\centering
		\includegraphics[width=\textwidth]{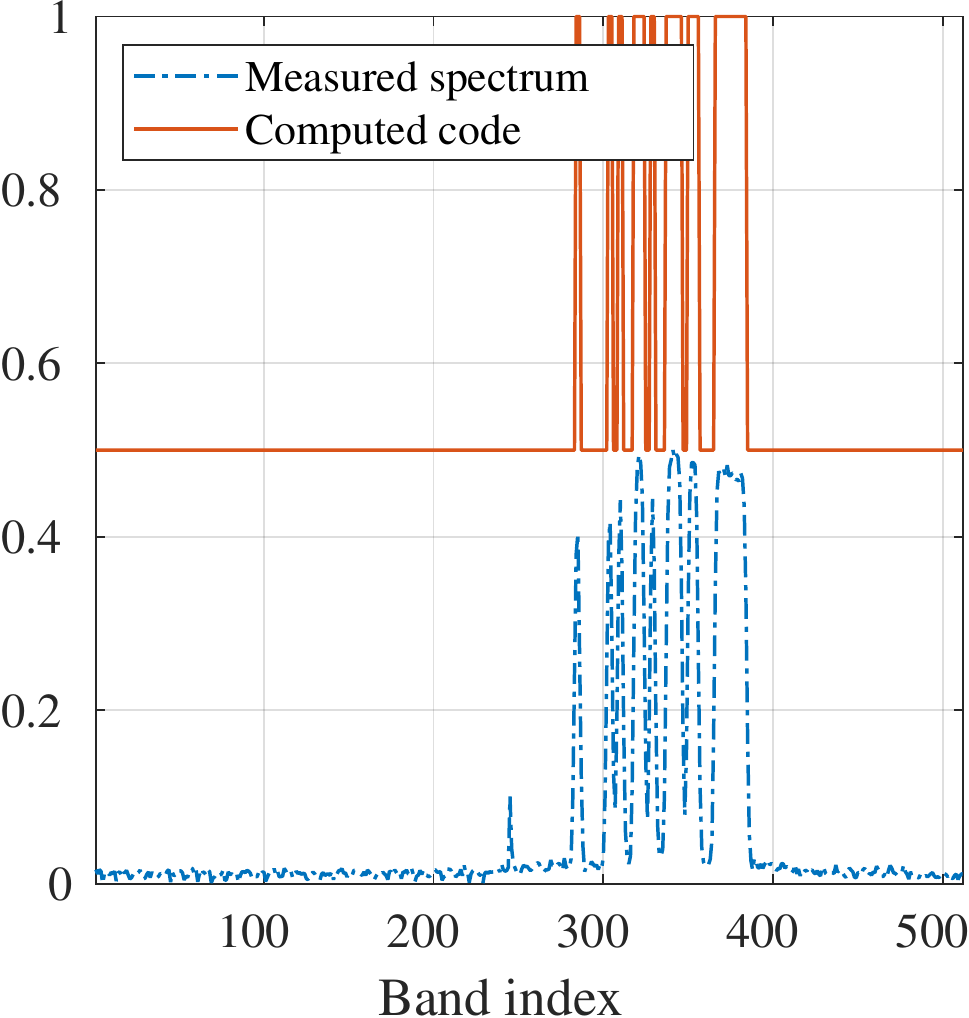}
		\caption{Code calibration}
	\end{subfigure}
	\hspace{0.05em}
	\begin{subfigure}[t]{0.45\columnwidth}
		\centering
		\includegraphics[width=\textwidth]{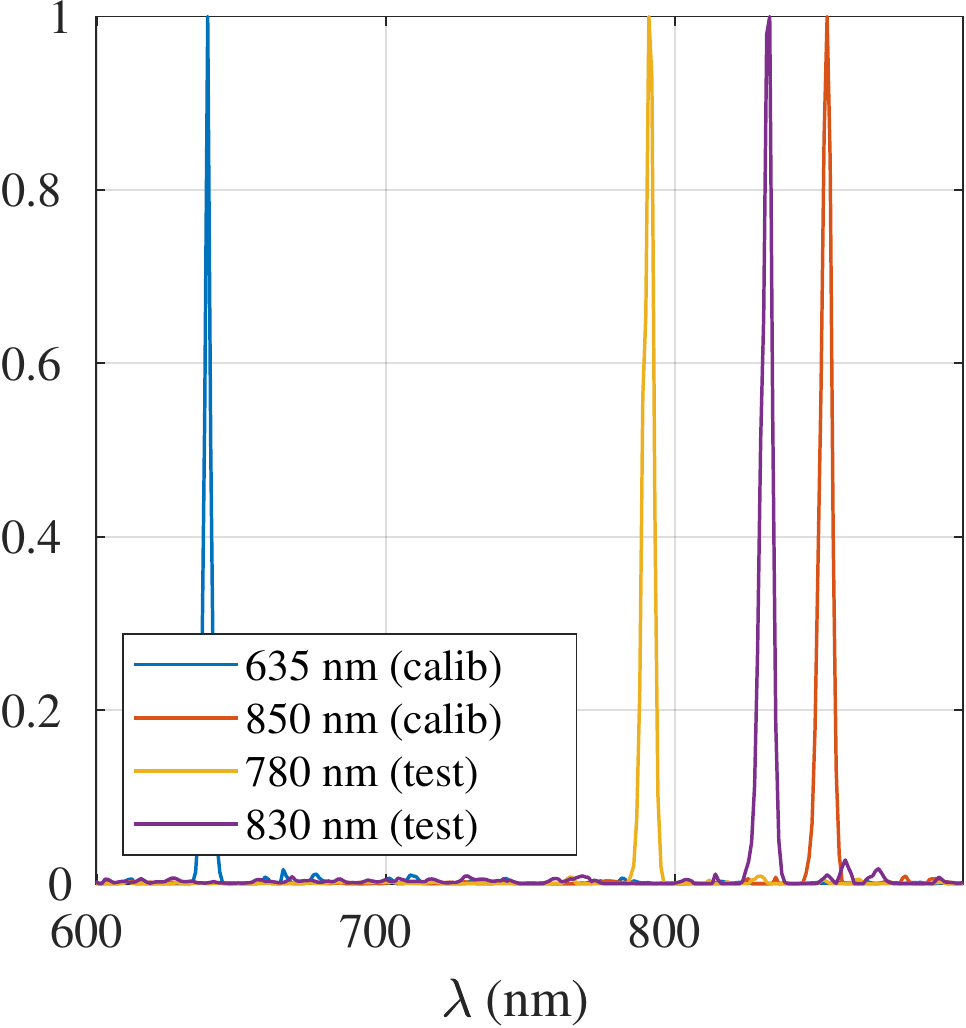}
		\caption{Wavelength calibration}
	\end{subfigure}
	\caption{\textbf{Wavelength calibration.} We first estimate the blur due to coded aperture by capturing a scene illuminated by a narrowband light source (635nm laser), giving us the code in (a). We then calibrate the correspondence between band index and wavelengths by capturing two separate scenes illuminated by known laser light sources (635nm, 850nm). The results of the two calibration are show in (b), where we capture two more scenes with 780nm and 830nm laser.}
	\label{fig:wvl_calib}
\end{figure}

\begin{figure}[!tt]
	\centering
	\begin{subfigure}[t]{\columnwidth}
		\centering
		\includegraphics[width=\textwidth]{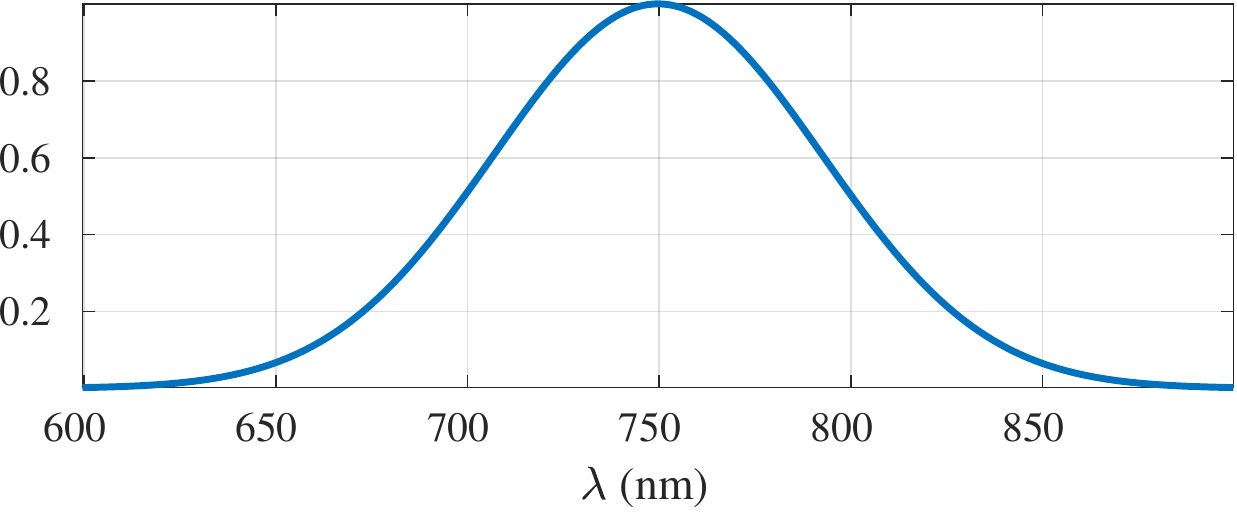}
		\caption{Target profile}
	\end{subfigure}
	\hspace{0.05em}
	\begin{subfigure}[t]{0.48\columnwidth}
		\centering
		\includegraphics[width=\textwidth]{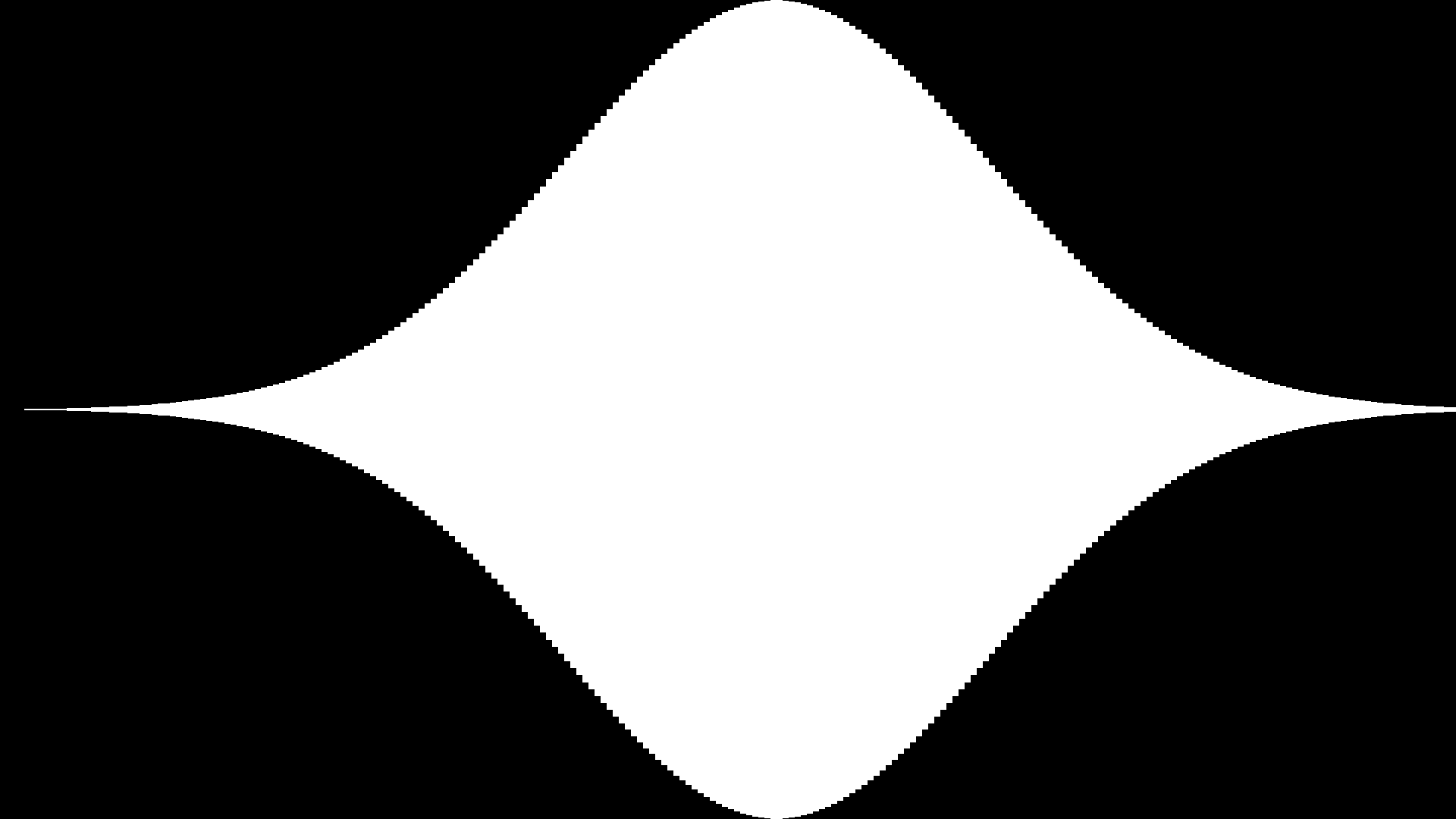}
		\caption{Without correction}
	\end{subfigure}
	\begin{subfigure}[t]{0.48\columnwidth}
		\centering
		\includegraphics[width=\textwidth]{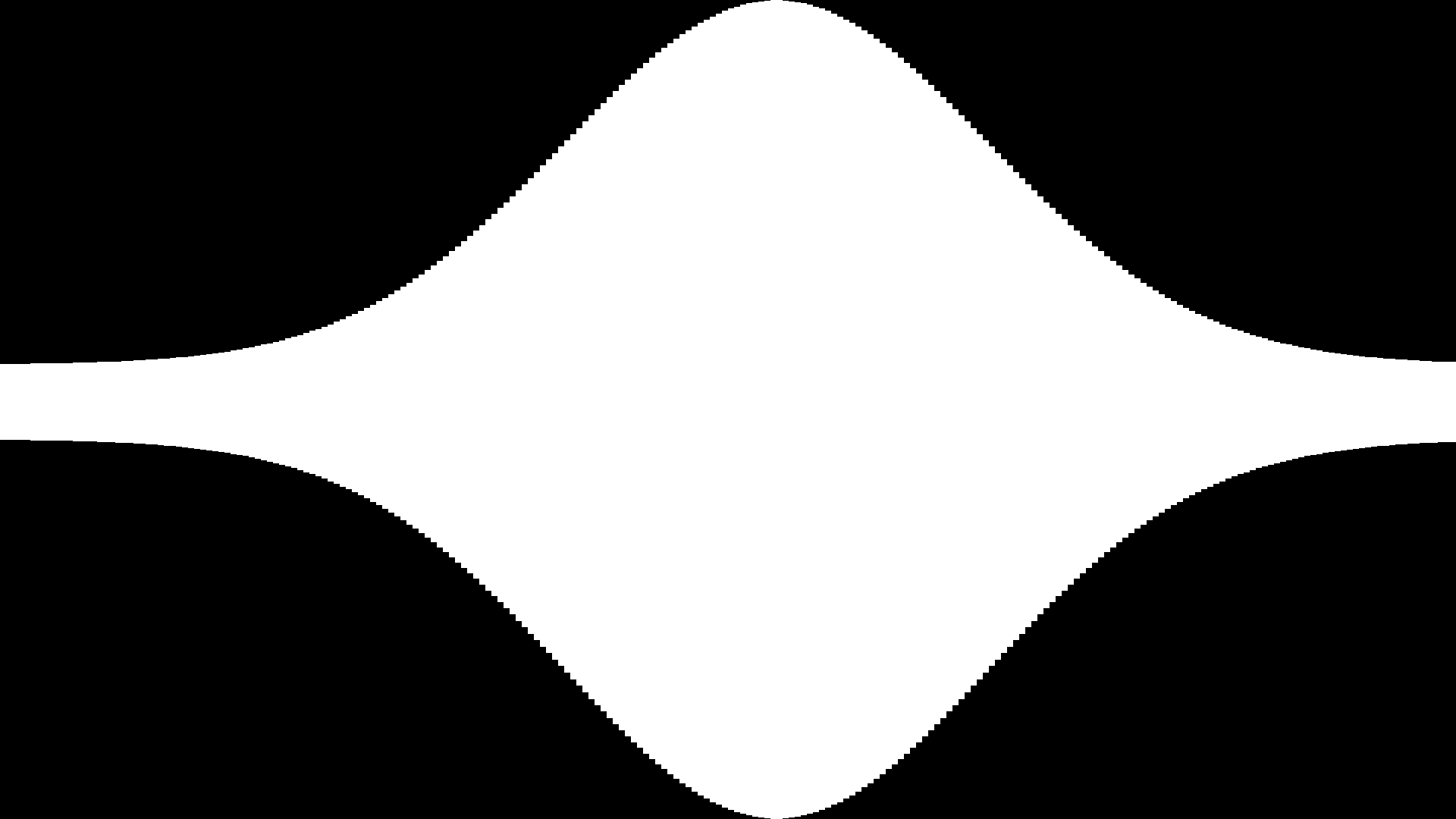}
		\caption{With correction}
	\end{subfigure}
	\caption{\textbf{Displaying desired spectral profile.} Given a target profile (a), we display a binary image on the SLM, as shown in (b), which ensures grayscale modulation despite wavelength-dependent gamma curve. However, since the SLM is $2f$ away from the camera sensor, there will be effects of diffraction. We counter this by adding a small DC offset, as shown in (c).}
	\label{fig:radiometric}
\end{figure}

\paragraph{Hardware prototype.} Figure \ref{fig:full_setup} shows a picture of our lab prototype with names of major components. 
The last lens in the setup was replaced by a $50$mm objective lens focused at infinity.
This led to a better spatial resolution than an achromat.

\paragraph{Calibration.}
As described in the main paper, our setup required calibration of coded aperture, wavelengths and spatial PSF. We detail the calibration procedure here.

\begin{enumerate}
	\item \emph{Coded aperture calibration}: This is required to capture the code that blurs the spectrum. We measure the coded aperture by illuminating a spectrally flat object (such as spectralon) with a laser of known wavelength and scanning the complete HSI. We then average all spatial pixels to get the spectrum of the scene. Since a laser can be treated as a discrete delta, the measured spectrum will be the coded aperture. We threshold the measured spectrum appropriately to get the binary coded aperture, as shown in Fig. \ref{fig:wvl_calib} (a).
	\item \emph{Wavelength calibration}: To find the correspondence between band index (1 - 256) and the corresponding wavelengths, we capture two scenes, each one comprised of a spectrally flat object illuminated by a narrowband laser light source. The averaged spectrum of the HSI is a blurred version of the laser spectrum. By deconvolving with the previously estimated coded aperture, we get location of the laser in terms of band index. We use this information along with laser wavelength to calibrate the correspondence.
	
	\item \emph{Spatial PSF}: To find the spatial blur kernel, we capture a single image of a $10\mu$m pinhole. Since the PSF is well conditioned, deblurring the spatial images is well conditioned.
	
	\item \emph{Radiometric calibration of SLM}: The LCoS SLM in our optical setup is based on twisted-nematic design, and hence has different gamma curves for different wavelengths. Since the spectrum on the SLM is a blurred version of the true spectrum, we cannot perform a column-wise gamma correction. Instead, we use the SLM only as a binary modulator and achieve grayscale modulation by varying height of each column as shown in Fig \ref{fig:radiometric} (b). This way, the SLM has a linear gamma curve for all wavelengths.
\end{enumerate}

\begin{figure}[!tt]
	\centering
	\begin{subfigure}[t]{0.48\columnwidth}
		\centering
		\includegraphics[width=\textwidth]{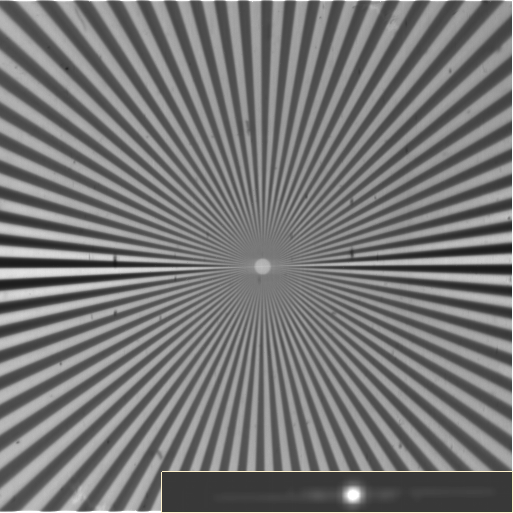}
		\caption{Raw}
	\end{subfigure}
	\hspace{0.05em}
	\begin{subfigure}[t]{0.48\columnwidth}
		\centering
		\includegraphics[width=\textwidth]{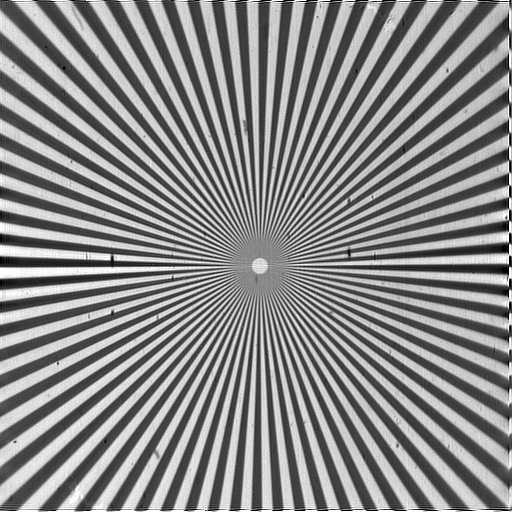}
		\caption{Deconvolved}
	\end{subfigure}
	\\
	\begin{subfigure}[t]{\columnwidth}
		\centering
		\includegraphics[width=\textwidth]{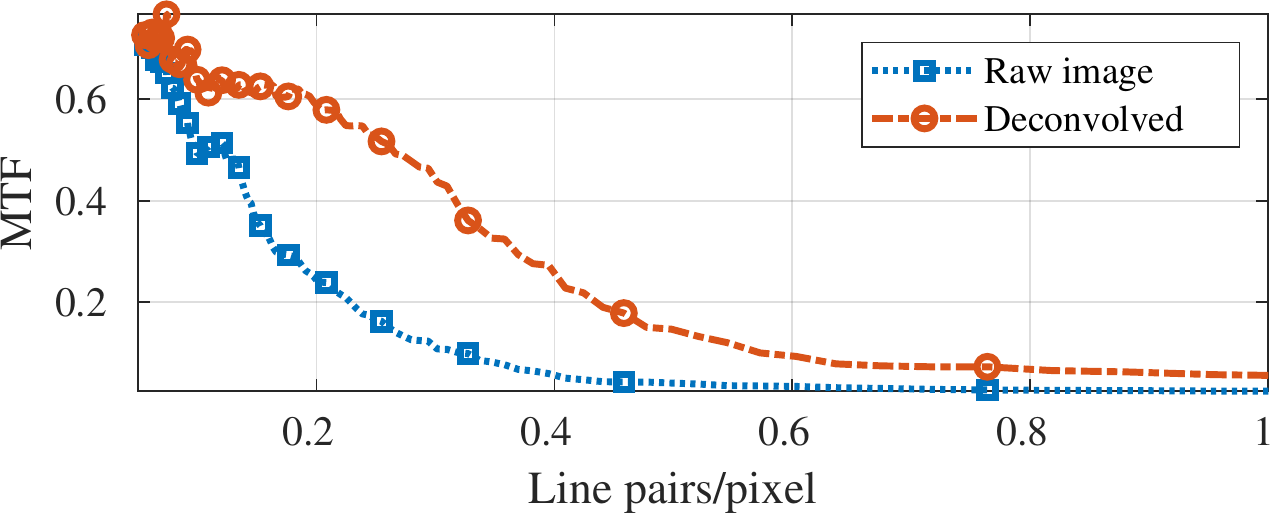}
		\caption{MTF}
	\end{subfigure}
	\caption{\textbf{Spatial deconvolution.} Due to design of an invertible spatial blur, the optical setup is capable of high resolution after deconvolution. (a) shows a raw image, with enlarged PSF in inset, (b) shows result of wiener deconvolution, and (c) shows a comparison of modulation transfer function (MTF). There is a marked increase in resolution both quantitatively and qualitatively.}
	\label{fig:mtf}
\end{figure}

\paragraph{Figures of merit.}
Our setup is capable of achieving spectral resolution of up to $3.3$nm over the wavelength range of $600 - 900$nm, which is the designed resolution (see \verb|KRISM.pdf| for further details).
Due to invertible spatial blur, our setup is capable of high resolution after deconvolution.
Figure \ref{fig:mtf} visualizes the captured image in (a) and deconvolved image in (b) of a sector star target. (c) shows plot of Modulation Transfer Function (MTF) as a function of line pairs per pixel.
Image was deconvolved using simple Wiener deconvolution.
The MTF30 after deconvolution was $0.45$ linepairs/pixel.

\paragraph{Handling diffraction due to SLM.}
Since the SLM is placed $2f$ away from the image plane, any pattern displayed on SLM will lead to a diffraction blur. To counter this effect, we always display ones in the middle of the pattern to be displayed on the SLM. This reduces the effect of diffraction while adding a simple offset to the data, which can be removed by capturing image with only the central part open.

\end{appendices}

{\small
\bibliographystyle{ieee}
\bibliography{refs}
}

\end{document}